\documentclass[11pt,letterpaper,twoside,final,english]{article}

\usepackage[dvipdfm,letterpaper]{geometry}
\usepackage{graphics}
\usepackage{epsfig}
\usepackage{ulem}
\usepackage{color}
\usepackage[breaklinks]{hyperref}
\hypersetup{colorlinks,citecolor=red}
\usepackage{titlesec}
\usepackage{setspace}
\usepackage{amsmath}
\usepackage{multirow}
\usepackage[round]{natbib}
\usepackage{mparhack}
\usepackage[table]{xcolor}   
\usepackage{booktabs}
\usepackage[bf]{caption}
\titleformat{\section}{\large\bfseries}{\thesection}{1em}{}
\titleformat{\subsection}{\bfseries}{\thesubsection}{1em}{}
\begin{document}
\onehalfspacing
\begin{titlepage}

\begin{center}
{\Large Model Risk in Real Option Valuation}\\ [10pt]

Carol Alexander$^a$ and Xi Chen$^b$\\ 
\end{center}

{\bf Abstract}

\onehalfspacing
\vspace{-.1cm}
\begin{quote}
\small \noindent 
We introduce a general decision-tree framework to value an option to invest/divest in a project, focusing on the model risk inherent in the assumptions made by standard real option valuation methods. We examine how real option values depend on the dynamics of project value and investment costs, the frequency of exercise opportunities, the size of the project relative to initial wealth,  the investor's risk tolerance (and how it changes with wealth) and several other choices about model structure.  For instance, contrary to stylised facts from previous literature, real option values can actually decrease with the volatility of the underlying project value and increase with investment costs. And large projects can be more or less attractive than small projects (ceteris paribus) depending on the risk tolerance of the investor, how this changes with wealth, and the structure of  costs to invest in the project. 
\\

[-10pt]


\end{quote}\vspace{-8pt}
\large\begin{tabular}{p{15.5cm}}
\toprule
~\\
\end{tabular}\normalsize

\noindent $^a$ Professor of Finance, University of Sussex, UK.
Email: c.alexander@sussex.ac.uk\\
\noindent $^b$ Research and Development Analyst, Oxford Risk, London. \\


\end{titlepage}
\onehalfspacing
\vspace{-20pt}
\setcounter{page}{1}
\section{Introduction}\label{sec.Intro}
\vspace{-10pt}
The original definition of a real option, first stated by  \cite{M1977}, is a decision opportunity for a corporation or an individual. It is a right, rather than an obligation, whose value is contingent on the uncertain price(s) of some underlying asset(s) and the costs incurred by exercising the option. The term `investment' real option concerns the opportunity to buy or sell a project (such as a property -- real estate, a company, a patent etc.) or a production process (such as an energy plant, or pharmaceutical research and development). Following \cite{KT1995}, we identify the \textit{project value} with a \textit{market price} that is the `break-even' price for which a representative decision maker would be indifferent between buying or not buying the project. Any higher  price would exceed her  value and she would not buy the project; any lower price would induce her to certainly  buy the project.\footnote{We use the term `market price' because this decision-maker's value is representative of the market as a whole.} Similarly, we identify the \textit{option strike} with the \textit{investment cost} for the project and in the following we shall use these two terms synonymously. The real option value (ROV) is the value of this decision opportunity to buy or sell the project; it is specific to the decision maker (i.e. the investor) and it will depend on her attitude to risk.  The ROV represents the \textit{certain} dollar amount, net of financing costs, that the decision maker should receive to obtain the same utility as the (risky) investment in the project. So, unlike the premium on a financial option, the ROV has no absolute accounting value.  In this setting, ROVs merely allow the subjective ranking of opportunities to invest in alternative projects.\footnote{For instance, just as financial investments are ranked using risk-adjusted performance measures, a pharmaceutical firm may compare the values of developing alternative products, or an exploration company may compare the values of drilling in different locations, or a property development corporation may compare the values of buying land in different locations. The same  utility function should be applied to value every investment opportunity (this characterises the decision maker) but the decision-maker's subjective views about the evolution of  future uncertainties (such as the value of a project and the associated costs of investment) is specific to each project. } 

The basic model for analysing  investment real options assumes that the project value follows a geometric Brownian motion (GBM) over a finite investment horizon, with a fixed or pre-determined investment cost (i.e. the option strike), where all risks can be hedged so that risk-neutral valuation (RNV) applies, and that decisions to invest or divest (i.e. exercise the option) may be implemented at any time up to the investment horizon. Much recent literature on investment real option analysis attempts to augment this basic model in various ways, to add special features that address particular practical problems.\footnote{For example, in the renewable energy industry, technological developments now help small-scale production projects gain competitive advantages, especially with various support schemes encouraging their adoption, and several recent papers use real option methods to analyse the value of these developments.\citet*{DL2015} model the value of an option to invest in a small-size production project taking into account temporary suspensions; \citet*{BMF2012} model the optimal capacity choice of producing renewable energy under different support schemes; \citet*{FMGJ2015} value an option to invest in microgrid assets, taking into account a variety of risks as well as the possibility of temporary suspension of production.} 

However, our paper addresses a much more fundamental issue. Instead of following the main strand of the real options literature, which adapts and extends the mathematical models to more closely reflect the real-world characteristics of a particular investment opportunity, we shall drill-down into the real option model  itself to examine the model risk inherent in the assumptions that are made, focussing on real options models that are most commonly applied in the literature. 

The way that we can analyse model risk is by first building a very general model which encompasses the standard assumptions, such as risk-neutral valuation,  as special cases. In fact, we make very flexible assumptions about: market price of the project; investment costs; the frequency of exercise opportunities; the size of the project relative to the decision-maker's wealth;  the decision maker's utility function; and how her risk tolerance changes with wealth. To our knowledge this is the first study of  model risk associated with these factors, some of which have not even been considered before.  

We do not assume that risks can be hedged by traded securities, but risk-neutral valuation techniques still apply in the special case of a linear utility function. In this case, \cite{G2011}  proves that the time-flexibility of the opportunity to invest in a project still carries a positive option value for a risk-averse decision maker, so that the paradigm of real options can still be applied to value a decision where none of the risks can be hedged. We also assume that the market price is the only stochastic factor and that the option strike may be -- but is not necessarily -- related to the market price of the project. 
In future work our general framework might be extendible to multiple correlated sources of uncertainty, where partial hedging is possible. But it is already a considerable challenge to build a model which is broad enough to encompass special cases corresponding to the specific choices of project characteristics  and investor preferences mentioned above.   Indeed, we already analyse a wide scope of questions about model risk, even without such model extensions. For instance: How is the ROV affected if the drift in market price is mean-reverting?  How does the frequency of exercise opportunities affect the ROV? How does the ranking of different opportunities change as we increase risk tolerance, or its sensitivity to wealth? Does the size of the investment relative to initial wealth effect the ROV, and if so how? And what is the effect of the fixed or pre-determined cost assumption relative to more general assumptions that costs are stochastic? 

In the following: Section \ref{sec.review} motivates our framework with a concise overview of the literature on investment real option models; Section \ref{sec.model} specifies the model mathematically;  Section \ref{sec.properties} answers the questions posed above assuming a GBM process for the market price of the project and Section \ref{sec.process} examines the sensitivity of ROVs to changing the price-process assumptions; Section \ref{sec.conclusion} summarises and concludes. An Appendix aids understanding of our  framework  with  some illustrative examples.

\section{Real Options to Invest in a Project}\label{sec.review}
\vspace{-10pt}

Most of the early literature on investment real options  focuses on opportunities to enter a tradable contract with pre-determined strike in a complete market. For early models including a stochastic strike, see \cite{MS1986}, \cite{Q1993} and \cite{K1998}. 
That is, the real option's pay-off can be replicated using tradable assets, so that all risks are hedgeable. Also, the decision opportunity can be exercised continuously at any time over the decision horizon as, for instance, in \cite{K1991}, \cite{G1996}, \cite{SM1998} and \citet*{PPS2005}. In this setting the option has the same value to all investors and so is priced as if the investor is risk neutral using  standard RNV techniques for American options. Basic real option models also suppose the forward  price follows a geometric Brownian motion (GBM) with total return equal to the risk-free rate, and the ROV is derived from a linear utility where exercise at a pre-determined strike may be taken at any point in time. \cite{CS1991}, \cite{T1993}, \cite{BK2000}, \cite{B2000}, \cite{YQ2002},  and numerous others since all employ this basic framework.

However, in practice, many investment opportunities encompassed by the original definition of \cite{M1977} are not standardised, tradable securities; their risks are only partially hedgeable, if at all;  investors are typically risk-averse; and investment costs are unlikely to be  pre-determined. So the RNV assumptions  are clearly inappropriate for many real options, particularly those in real estate, research \& development or company  mergers \& acquisitions. In many applications the  underlying  market is incomplete and the project is not something that can be bought and sold in a market, indeed the transacted price may be negotiated between individual buyers and sellers. In this setting there is no unique value for a real option -- 
it will be  specific to the decision maker, depending upon her subjective views about the costs and benefits of investment in the project, and upon her risk preferences. For instance, if an oil exploration company must decide whether to drill in location A or location B, their views about the benefits of drilling in each location will depend on their subjective beliefs about the market price of oil in the future as well as their risk tolerance. Hence, there is no single accepted definition in the real option literature for the stochastic price process which drives the project value. For instance,  \citet*{S2001,MT2002,AP2011,CRS2011, AMS2012,JDSZ2013} associate the market price of the project as the value of the project's outputs; \citet*{AS2004,BFJLR2008,F2008,F2015,MCP2018} model the demand/production capacity and their market price is the value of the products that generate the project’s profit stream; and \citet*{MCCC2011,LWY2016} use the net present value of the project's future  cash flows.

Several papers consider risk-averse decision makers in incomplete or partially complete markets, often for decision makers with a utility function that admits closed-form solutions via maximization of an expected utility.  \cite{H2007}, \cite{MW2007},  \cite{G2011}, \cite{AP2013} and \citet*{HHKN2016} all investigate continuous-time, deterministic-strike real options using a two-factor GBM framework in which the value of the project to the investor is stochastic and possibly correlated with the price of a liquidly traded asset that may be used to hedge the investment risk. In common with our approach, which is described in detail in the next section,  price processes are discounted to present value terms so that a univariate time $0$ utility function may be applied to maximise the expected utility of the maturity pay-off.

The process of real option valuation also determines an optimal exercise, assuming that this may occur at any time in a continuous future. In practice these opportunities may only be possible at fixed discrete points in time, such as at a weekly meetings of the investment committee or monthly meetings of the board of directors, but there is no research to date on the effect of infrequent exercise opportunities on the ROV.  Our framework allows a decision-maker to quantify this effect. For instance, the optimal decision at time $0$ may be to invest if the market price falls by a given amount. But  if the board  controls the decision to exercise the option, and the board meets only once per month or even less frequently, then the  ROV is likely to be lower than it is when exercise can happen at any point in time. How much lower? This question can be addressed by designating a proper sub-set of nodes as decision nodes in our general model.  

\cite{T1996}, \cite{S1996},  \cite{BD2005}, \citet*{BDH200502,BDH200501} and \cite{Smith2005} all employ decision trees but, unlike us, they assume market-priced risk and thus adopt the RNV framework, avoiding the explicit use of a utility.  \cite{MM1985}, \cite{TM1987}, \citet*{RDV2008}  and \citet*{MCCC2011} also use decision trees, but with a constant risk-adjusted discount rate. Only \cite{G2011} applies a utility function in a decision tree for real option valuation and the exponential (CARA) utility that he applies has substantial limitations -- see Section \ref{sec.utility} for details.

\section{A General Framework for Investment Real Option Valuation}\label{sec.model}
\vspace{-10pt}
We apply the general hyperbolic absolute risk aversion (HARA) utility, introduced by \cite{M1968} and \cite{M1971}. The downside of this generality is that analytic solutions are only available in special cases (e.g. with an exponential utility and GBM). We do not rely on expansion approximations, which can be unreliable. So all results in this paper are generated numerically. However, they can all be verified  -- and extended by changing the (numerous) model parameters -- using code that is available to download. 
 
When risks are unhedgeable the forward market price measure is subjective to the decision maker, with the risk-neutral measure arising as the special case of a linear utility and a risk-free expected return. The decision maker holds subjective views not only about the evolution of the market price but also about the stream of cash flows (if any) that would be realised if she  enters the investment or begins the project. In most applications cash flows would reflect the individual management style of the decision maker, e.g. aggressive, expansive, recessive etc. Our investment costs may have pre-determined and/or stochastic components. 

Exercise opportunities are discrete and are modelled using a binomial price tree with decision nodes, corresponding to the possibility for exercise, placed at every $k$ steps until a finite decision horizon $T$. The consequence of the decision (which is taken at time $0$) is valued at some finite investment horizon, $T^\prime > T$.\footnote{We do not equate these horizons with the maturity of the investment, i.e. cash flows from the project may continue beyond $T^{\prime}$, so the market price is not constrained to be the discounted expected value of the cash flows.} 
 Additionally, the decision maker is characterised by her initial wealth, $w_0$ which represents the current net worth of all her assets, and a HARA utility function $U(w)$ which reflects her risk tolerance $\lambda$ and how this changes with wealth.  
\vspace{-5pt}
\subsection{Market Prices and Cash Flows}\label{sec.prices}
\vspace{-5pt}
All future market prices and cash flows are expressed in time $0$ terms by discounting at the decision-maker's borrowing rate $r$.\footnote{This rate depends on the business risk of the project as perceived by the financer, not as perceived by the decision maker. It may also depend on the decision-maker's credit rating but this assumption is not common in the real options literature.}
Thus the investor borrows funds at rate $r$ to invest, rather than financing the cost from her initial wealth, which  is not available for investing in this project.\footnote{To avoid additional complexity we do not consider that projects could be financed from wealth, even though this would be rational if wealth is liquid and $r$ is greater than the return on wealth, $\tilde r$.} We suppose that $r$ is a constant, risk-free rate. First we allow the market price of the project to follow a GBM, as usual, so that the discounted forward market price $p_t$ evolves as:
\begin{equation}\label{eqn.gbm}
\frac{dp_t}{p_t} = (\mu - r)\, dt + \sigma \,dW_t, \quad \mbox{for} \quad 0 < t \le T^{\prime},\vspace{-.1cm}
\end{equation}
where $\mu$ and $\sigma$ are the decision-maker's subjective drift and volatility associated with $p_t$ and $W_t$ is a Wiener process.   
Then $p_t$ has a lognormal distribution,
$p_t \sim \log \mbox{N}\left((\mu - r)\, t, \sigma^2t\right)$. 

It is convenient to use a binomial tree discretisation of (\ref{eqn.gbm}) in which the price can move up or down by factors $u$ and $d$, so that $p_{t+1} = p_t u$ with probability $\pi$ and otherwise $p_{t+1} = p_t d$.
No less than eleven different binomial parameterisations for GBM are reviewed by \cite{C2008}. 
\cite{Smith2005}, \cite{BD2005}, \citet*{BDH200502,BDH200501}, \cite{SA1993} and others employ the `CRR' parameterisation of \citet*{CRR1979}. However, the \cite{JR1982} parameterisation, which is commonly used by option traders, is more stable for low levels of volatility and when there are only a few steps in the tree.  
Thus we set
\begin{equation}\label{eqn.jr}
m=\left[\mu - r -0.5\sigma^2\right]\Delta t, \quad u = e^{m+\sigma \sqrt{\Delta t}}, \quad d=e^{m-\sigma \sqrt{\Delta t}}\quad \mbox{and} \quad \pi=0.5.\vspace{-.1cm}
\end{equation}
 
Secondly, we employ a modification of (\ref{eqn.gbm}) that represents a `boom-bust' scenario via a regime-dependent process, which trends upward with a low volatility for a sustained period and downward with a high volatility for another sustained period. Thus we set:
\begin{equation}\label{eqn.bb}
\frac{dp_t}{p_t} =  \left\{ \begin{array}{rl}
\vspace{0.25cm}
(\mu_1 - r) dt + \sigma_1 dW_t, & \quad \mbox{for} \quad 0 < t \le T_1,\\

(\mu_2 - r) dt + \sigma_2 dW_t, & \quad \mbox{for} \quad T_1 < t \le T^{\prime}.\\
\end{array}  \right.\vspace{-.1cm}
\end{equation} 

Finally, we consider a case where the decision maker believes the market price  will mean-revert over a relatively short time horizon. To represent this we suppose the expected return decreases following a  price increase but increases following a price fall, as in the Ornstein-Uhlenbeck process:
\begin{equation}\label{eq.OU}
d\ln p_t=-\psi\ln\left(\frac{p_t}{\bar{p}}\right)dt+\sigma dW_t\vspace{-.1cm}
\end{equation}
where $\psi$ denotes the rate of mean reversion to a long-term price level $\bar{p}$. Following \cite{NR1990} (NR) we employ the following binomial tree parameterisation for the discretised Ornstein-Uhlenbeck process:
\begin{subequations}\label{eq.OUdiscrete}
\begin{equation}
u=e^{\sigma\sqrt{\Delta t}},\quad d=u^{-1}\vspace{-.1cm}
\end{equation}
\begin{equation}
\pi_{{\bf s}(t)}=\left\{ \begin{array}{lcc}\vspace{.3cm}
1,&\quad&0.5+\nu_{{\bf s}(t)}\sqrt{\Delta t}/2\sigma>1\\ \vspace{.3cm}
0.5+\nu_{{\bf s}(t)}\sqrt{\Delta t}/2\sigma,&&0\leq 0.5+\nu_{{\bf s}(t)}\sqrt{\Delta t}/2\sigma\leq 1\\
0, &&0.5+\nu_{{\bf s}(t)}\sqrt{\Delta t}/2\sigma<0
\end{array}\right.
\end{equation}
where
\begin{equation}
\nu_{{\bf s}(t)}=-\psi\ln\left(\frac{p_{{\bf s}(t)u}}{\bar p}\right)
\end{equation}
is the local drift of the log price process, and so the price process has local drift:
\begin{equation}\label{eq.NRmu}
\mu_{{\bf s}(t)}=-\psi\ln\left(\frac{p_{{\bf s}(t)u}}{\bar p}\right)+0.5\sigma^2+r.\vspace{-.1cm}
\end{equation}
\end{subequations}
Note that when $\psi=0$ there is a constant transition probability of 0.5 and the NR parameterisation is equivalent to the parameterisation (\ref{eqn.jr}) with $m=0$. 

The cash flows may depend on the market price of the project (as do revenues from pharmaceutical sales, for instance, or rents from a property). To capture this feature, we let ${{\bf s}(t)}$ denote the state of the market  at time $t$, i.e. a path of the market price from time $0$ to time $t$ 
 written as a string of $u$'s and $d$'s with $t$ elements, e.g. ${{\bf s}(3)}=uud$. Let CF$_{{\bf s}(t)}$ denote the cash flow in state ${{\bf s}(t)}$ at time $t$. 
Regarding cash flows as dividends we call the price excluding all cash flows before and at time $t$ the `ex-dividend' price, denoted $p^-_{{\bf s}(t)}$. 
At the time of a cash flow CF$_{{\bf s}(t)}$ the market price follows a path which jumps from  $p^-_{{\bf s}(t)}$ to $p^+_{{\bf s}(t)} = p^-_{{\bf s}(t)}+\mbox{CF}_{{\bf s}(t)}$. The investor does not receive $\mbox{CF}_{{\bf s}(t)}$ if she  invests in the project at time $t$, but she  does receives it if she  divests in the project at time $t$.\footnote{The alternative assumption that she  receives the cash flow at time $t$ only when she  invests is also possible. The subscript ${\bf s}(t)$ denotes a particular realisation of the random variable that carries the subscript $t$, e.g. $p^+_{uud}$ and $p^-_{uud}$ are the left and right limits of a realisation of $p_t$ when $t=3$. From henceforth we only use the subscript ${\bf s}(t)$ when it is necessary to specify the state of the market -- otherwise we simplify notation using the subscript $t$. Also, assuming cash flow is not received when divesting at time $t$, $p^-_t$ is a limit of $p_t$ from the right, not a limit from the left.}     
The dividend yield is defined as 
\begin{equation}\label{eqn.div}
\delta_{{\bf s}(t)} = \frac{p^+_{{\bf s}(t)} - p^-_{{\bf s}(t)}}{p^+_{{\bf s}(t)}}\,.\vspace{-.1cm}
\end{equation} 
 We assume that dividend yields are deterministic and time but not state dependent, using the simpler notation $\delta_t$. Then the cash flows are both time and state dependent.\footnote{If the cash flow is not state dependent, then the dividend yield must be state dependent. The state dependence of cash flows induces an autocorrelation in them because the market price is autocorrelated. For this reason, defining an additively separable multivariate utility over future cash flows as in \cite{SN1995} and \cite{SM1998} may be problematic. } 

\vspace{-5pt}
\subsection{Costs and Benefits of Investment and Divestment}\label{sec.termw}
\vspace{-5pt}
Assumptions about costs are critical determinants of real option values. For example, if the cost is stochastic and perfectly correlated with a martingale discounted project price then the RNV price will be zero, yet a risk-averse investor could still place a positive value on such a decision opportunity. We shall allow fixed, deterministic or stochastic investment costs which may be related to the market price, because this is relevant to many practical situations (e.g. purchasing land for development or investing in pharmaceutical research). Hence, the investment cost at time $t$, in time 0 terms, is:
\begin{equation}\label{eqn.K}
K_{t}=\alpha K+(1-\alpha)\,g(p_{t}^-), \quad 0\leq \alpha\leq 1,\vspace{-.1cm}
\end{equation}
where $K$ is a constant in time 0 terms and $g:\mathcal{R} \rightarrow \mathcal{R}$ is some real-valued function linking the investment cost to the market price. This function controls the correlation between costs and market price of the project. For instance, cost is perfectly correlated with price when $g$ is linear and increasing. When $\alpha = 1$ we have a standard real option with a pre-determined strike $K$, such as might be employed for oil exploration decisions. When $\alpha = 0$ there is only a variable cost, which is a deterministic function of the market price $p_{t}^-$.  This might be employed for real estate, research and development or merger and acquisition options. The intermediate case, with $0< \alpha< 1$ has an investment cost with both fixed and variable components.

We assume that initial wealth $w_0$ earns a constant, risk-free lending rate $\tilde r$, as do any cash flows paid out which are not re-invested. Any cash paid into the project (e.g. development costs) is financed at the borrowing rate $r$. The financial benefit to the decision-maker on investing (action I) at time $t$ is the sum of any cash flows paid out and not re-invested plus the terminal market price of the project. 
Thus the wealth of the decision-maker at $T^\prime$, in time $0$ terms, following investment at time $t$ is
\begin{equation}\label{eqn.CF}
w_{t,T^\prime}^I=e^{(\tilde r-r)T^\prime} w_0+\sum^{T^\prime}_{s=t+1}e^{(\tilde r-r)(T^{\prime}-s)}\mbox{CF}_s+ p_{T^\prime}^- - K_t.
\vspace{-0.1cm}
\end{equation}

Some projects pay no cash flows, or any cash flows paid out are re-invested in the project. Then the financial benefit of investing at time $t$, in time $0$ terms, is simply  the cum-dividend price $\widetilde{p}_{t,{T^\prime}}$ of the project accruing from time $t$. If a decision to invest is made at time $t$, with $0 \le t \le T$, then $\widetilde{p}_{t,t} =p_t$ but the evolution of $\widetilde{p}_{t,s}$ for $t < s \le T^\prime$ differs to that of $p_s$ because $\widetilde{p}_{t,s}$ will gradually accumulate all future cash flows from time $t$ onwards. In this case, when the decision maker chooses to invest at time $t$ her wealth at time $T^\prime$ in time 0 terms is
\begin{equation}\label{eqn.CDP}
\widetilde{w}_{t,{T^\prime}}^I = e^{(\tilde r-r)T^\prime}w_0  + \widetilde{p}_{t,{T^\prime}}-K_t.\vspace{-.1cm}
\end{equation}
Note that if $\tilde r = r$ then also $\sum^{T^\prime}_{s=t+1}e^{(\tilde r-r)(T^{\prime}-s)}\mbox{CF}_s+ p_{T^\prime}^- = \tilde{p}_{t,T^\prime}$ in (\ref{eqn.CF}). 

Similarly, if the decision maker already owns the project at time $0$ and chooses to divest (action S) at time $t$, the time $0$ value of her wealth at time $T^\prime$ is 
\begin{equation}\label{eqn.CFdis}
w_{t,{T^\prime}}^S = e^{(\tilde r-r)T^\prime}w_0 + \sum_{s=1}^{t-1}e^{(\tilde r-r)(T^{\prime}-s)}\mbox{CF}_s+ p^+_{t}- p_0 .\vspace{-.1cm}
\end{equation} 
Note that $p_0$ is subtracted here because we assume the investor has borrowed funds to invest in the project. Alternatively, if there are no cash flows, or they are re-invested,
\begin{equation}\label{eqn.CDPdis}
\widetilde{w}_{t,{T^\prime}}^S = e^{(\tilde r-r)T^\prime} w_0+ \widetilde{p}_{0,t} -p_0 .\vspace{-.1cm}
\end{equation}
Finally, the wealth  $w_{t,T^\prime}^D$ resulting from a defer decision at time $t$ depends on whether she  invests (or divests) later on. This must therefore be computed using backward induction as described in the next sub-section.
\vspace{-5pt}
\subsection{Optimal Decisions and Real Option Value}\label{sec.decision}
\vspace{-5pt}
Denote the decision-maker's utility function by $U: {\cal R} \rightarrow {\cal R}$. As in any decision problem, we shall compare the expected utility of the outcomes resulting from investment with the utility of a base-case alternative, which in this case is to do nothing (so  terminal wealth remains at $w_0$ in time 0 terms).\footnote{For brevity, we only describe the backward induction step for the decision to invest but it is similar for the decision to divest. That is, for the divestment decision we compare the expected utility of the outcomes resulting from remaining invested with the utility of the alternative, to divest.}  
The option to invest at time $t$ has time $0$ utility value $U_{t,{T^\prime}}^I =  U\left(w_{t,{T^\prime}}^I \right)$, but 
since $w_{t,{T^\prime}}^I$ is random so is $U_{t,{T^\prime}}^I$, and we use the expected utility $$\mbox{E}\left[U_{{\bf s}(t),{T^\prime}}^I\right] = \mbox{E}\left[U\left(w_{{\bf s}(t),{T^\prime}}^I \right)\right],$$ as a point estimate. Then, given a specific decision node at time $t$, say  when the market is in state ${\bf s}(t)$, the potential investor chooses to invest if and only if $$\mbox{E}\left[U_{{\bf s}(t),{T^\prime}}^I\right] > \mbox{E}\left[U_{{\bf s}(t),{T^\prime}}^D\right],$$ 
%
and we set 
\begin{equation}\label{eqn.maxEU}
\mbox{E}\left[U_{{\bf s}(t),{T^\prime}}\right] = \mbox{max} \left\{\mbox{E}\left[U_{{\bf s}(t),{T^\prime}}^I\right] , \mbox{E}\left[U_{{\bf s}(t),{T^\prime}}^D\right]\right\}. \vspace{-.1cm}
\end{equation}
Since there are no further decisions following a decision to invest, $\mbox{E}\left[U_{{\bf s}(t),{T^\prime}}^I\right]$ can be evaluated directly, using the utility of the terminal wealth values obtainable from state ${\bf s}(t)$ and their associated probabilities. However, $\mbox{E}\left[U_{{\bf s}(t),{T^\prime}}^D\right]$ depends on whether it is optimal to invest or defer at the decision nodes at time $t+1$. Thus, the expected utilities at each decision node must be computed via backward induction. 

First we evaluate (\ref{eqn.maxEU}) at the last decision nodes in the tree, which are at the time $T$ that option expires. These nodes are available only if the investor has deferred at every node up to this point. We associate each ultimate decision node with the maximum value (\ref{eqn.maxEU}) and select the corresponding optimal action, $I$ or $D$. 
Now select a penultimate decision node; say it is at time $T-k\Delta t$. If we use a recombining binomial tree to model the market price evolution, it has $2^k$ successor decision nodes at time $T$.\footnote{The recombining assumption simplifies the computation of expected utilities at the backward induction step. However, we do not require that the binomial tree is recombining so the number of decision nodes could proliferate as we advance through the tree. Note that the state price tree will recombine if cash flows are determined by a time-varying but not state-varying dividend yield.} 
Each market state ${\bf s}(T-k\Delta t)$ has an associated decision node. Each one of its successor nodes is at a market state ${\bf s}^*(T)$ that is attainable from state ${\bf s}(T-k\Delta t)$, and has an associated probability $\pi_{{\bf s}^*(T)}$ determined by the state transition probability of $0.5$, given that we employ the parameterisation (\ref{eqn.jr}).\footnote{So if the tree recombines these probabilities are $0.5^k, k0.5^k, k!/(2!(k-2)!)0.5^k, ...., k0.5^k, 0.5^k$ under the JR parameterization (\ref{eqn.jr}). If the CRR parameterisation is employed instead the transition probabilities for a recombining tree would be more general binomial probabilities.} Using the expected utility associated with each attainable successor node, and their associated probabilities, we compute the expected utility of the decision to defer at time $T-k\Delta t$. More generally, assuming decision nodes occur at regular time intervals, the backward induction step is: 
\begin{equation}\label{eqn.backind}
\mbox{E}\left[U_{{\bf s}(t-k\Delta t),{T^\prime}}^D\right] = \sum_{{\bf s}^*(t)} \pi_{{\bf s}^*(t)}\mbox{E}\left[U_{{\bf s}^*(t),{T^\prime}}\right], \quad t = k\Delta t, 2k\Delta t, \ldots,T-k\Delta t, T. \vspace{-.1cm}
\end{equation}
At each decision node we compute (\ref{eqn.backind}) and associate the node with the optimal action and its corresponding maximum expected utility. We repeat the backward induction until we arrive at a single expected utility value associated with the node at time $0$. Finally, the option value is the certainty equivalent (CE) of this expected utility, less the wealth resulting from the base-case alternative, i.e. $w_0$, where CE$(w) = U^{-1}\left(\mbox{E}[U(w)]\right)$ for any monotonic increasing utility $U$. The minimum value of zero applies when the project would never be attractive whatever its future market price. 
\vspace{-5pt}
\subsection{Risk Preferences}\label{sec.utility}
\vspace{-5pt}
 Much previous research on decision analysis of real options, reviewed in Section \ref{sec.review}, employs either risk-neutrality or an exponential utility function, which may be written in the form 
\begin{equation}\label{eqn.exp}
U(w) = -\lambda \exp\left(-\frac{w}{\lambda}\right), \vspace{-.1cm}
\end{equation}
where $w$ denotes the terminal (time $T^{\prime}$) wealth of the decision maker, expressed in time $0$ terms and $\lambda > 0$ denotes her risk tolerance and $\gamma = \lambda^{-1}$ is her risk aversion. Note that $w$ is a random variable taking values determined by the decision-maker's (subjective) views on the evolution of the market price and the decisions she  takes before time $T^{\prime}$.  Under (\ref{eqn.exp}) we have
\begin{equation}\label{eqn.expce}
\mbox{CE}(w)=-\lambda \log \left(-\frac{\mbox{E}[U(w)]}{\lambda}\right).\vspace{-.1cm}
\end{equation}
This function is frequently employed because it has special properties that make it particularly tractable \citep*[see][Chapter 6]{DKLS2006}.   
\begin{enumerate}
\item The exponential function (\ref{eqn.exp}) is the only utility with a CE that is independent of the decision-maker's  initial wealth, $w_0$.\footnote{In other words, adding a constant to $w$ results in only in an affine transformation which does not change the form of the utility. This is also known as the `delta' property. It implies that we obtain the same option value, and the same optimal decisions, whether $w$ denotes the P\&L or net wealth.}
\item The Arrow-Pratt coefficient of risk aversion is constant, as $-U^{\prime \prime}(w)/U^{\prime}(w) = \gamma$. Thus, the exponential utility (\ref{eqn.exp}) represents decision makers with constant absolute risk aversion (CARA) and $\lambda$ in (\ref{eqn.exp}) is the absolute risk tolerance.
\item The CE of an exponential utility is additive over independent risks.
When $x_t \sim$NID$(\mu,\sigma^2)$ and $w_T = x_1 + \ldots + x_T$ then $\mbox{CE}(w_T) = \mu T - (2\lambda)^{-1}\sigma^2T.$ 
\end{enumerate}
\noindent Unfortunately, properties 1 and 2 are very restricting. A major critique is that CARA decision makers leave unchanged the dollar amount allocated to a risky investment when their initial wealth changes; indeed, the decision maker's wealth has no influence on her valuation of the option.   
Property 3 implies that when cash flows are normally and independently distributed (NID) the decision-maker's risk premium for the sum of cash flows at time $t$ is $(2\lambda)^{-1}\sigma^2t$, so it scales with time at rate $(2\lambda)^{-1}\sigma^2$. This could be used to derive the risk-adjustment term that is commonly applied to DCF models and in the influential book by \citet*{CKM1990}.\footnote{Setting $\mu \exp[r^a] = \mu - (2\lambda)^{-1}\sigma^2$ gives $-r^a = \log[1 -(2\lambda\mu)^{-1}\sigma^2]$, so $r^a \approx (2\lambda\mu)^{-1}\sigma^2$.}  

Economic analysis is commonly based on an inter-temporal consumption framework, for instance as in \cite{SN1995} and \cite{SM1998}. Here decision-marker's have a multivariate utility given by a sum of time-homogenous exponential utilities over serially-independent future cash flows. These particular assumptions are necessary  because only then is the CE additive over independent random variables, and the ROV is obtained by backward induction on the CE. But in finance it is standard to base all utility on final wealth with future values discounted to time 0 terms, as in \cite{H2007}, \cite{MW2007} and \cite{G2011} and many others since. 

This discounting is an essential element of our approach because it allows the backward induction step to be defined on the expected utility relative to \textit{any} HARA utility which is just a univariate function of terminal wealth. It is very important to note that using exponential preferences over independent cash flows in this framework will yield the same decision (to invest, or to defer)  at \textit{every node in the tree}  because the uncertainties faced at any time are just  scaled versions of the initial uncertainties. 
Therefore, to reach beyond the restricting and trivial exponential-utility solutions, we shall assume decision makers are endowed with a HARA utility, even though this comes  at the expense  of analytic tractability, which may be written in the form:
\begin{equation}\label{eqn.HARA}
U(w)= - (1-\eta)^{-1}[1+\frac{\eta}{\lambda w_0}(w-w_0)]^{1-\eta^{-1}}, \ \ \mbox{for}\ w > (1-\eta^{-1}\lambda)w_0.\vspace{-.1cm}
\end{equation}
where $\lambda$ denotes the \textit{local} relative risk tolerance, which increases linearly with wealth at the rate $\eta$.\footnote{Note that an investor's risk tolerance and its sensitivity to wealth may be specified  using the techniques introduced by \cite{KR1993}. Relative risk tolerance is expressed as a percentage of wealth, not in dollar terms. So if, say, $\lambda=0.4$ the decision maker is willing to take a gamble with approximately equal probability of winning 40\% or losing 20\% of her wealth, but she  would not bet on a 50:50 chance (approximately) of winning $x\%$ or losing $x\%/2$ for any $x > 0.4$.} Here both absolute and relative risk aversion can increase with wealth, depending on the value of $\eta$.
We consider four special cases: when $\eta \rightarrow 0$ we converge to the exponential utility; $\eta = 1$ corresponds to the displaced logarithmic utility;  $\eta = 0.5$ gives the hyperbolic utility; and $\eta = \lambda$ yields the power utility.

\section{Model Risk under GBM}\label{sec.properties}
\vspace{-10pt}
 
Investment real option values, and hence the ranking of alternatives, are influenced by a large number of variables: (i) the fixed/variable cost structure of the investment; (ii) the scheduling of exercise opportunities; (iii) the investor's risk tolerance and its sensitivity to wealth; (iv) the size of the project relative to the initial wealth; and (v) the decision-maker's views on the evolution of market prices.  In this section we first investigate how the ROV depends upon  assumptions (i) and (ii) when decision makers have exponential utilities and the market price follows a given GBM (\ref{eqn.gbm}).\footnote{From henceforth we set $\tilde r = r$. In the previous section it was important to specify the model in all its generality, but  insights to model risk based on using different lending and borrowing rates is a less important problem which is therefore left to future research. To this end, the downloadable code accompanying this article allows one to set $\tilde r \ne r$.} Then we investigate (iii) by fixing the risk tolerance $\lambda$ and seeing how our assumption about the rate at which  $\lambda$ increases with wealth (which is zero in the exponential case) influences the ROV. Then, for property (iv), employing both exponential and logarithmic utilities for illustration, we show how the ROV changes with the size of the project revenues relative to initial wealth. Investigations of the ROV sensitivities to expected returns and volatility, and how the ROV behaves for market price processes other than GBM, are dealt with in Section \ref{sec.process}.

\vspace{-5pt}
\subsection{Investment Costs}\label{sec.costRN}
\vspace{-5pt}
A variety of investment cost assumptions are made in the literature.\footnote{Among many others: \citet*{CRS2011} and \citet*{AS2004} use fixed investment costs; deterministic costs are applied by \citet*{MCP2018}, \citet*{DL2015} and \citet*{F2015}; \citet*{CKS2017} and \citet*{MCCC2011} have stochastic investment costs; and \citet*{F2008} assumes a mixture of fixed and stochastic costs.} In some applications -- for instance, when a licence to drill for oil has been purchased, and the decision concerns whether the market price of oil is sufficient to warrant exploration -- a fixed-strike or pre-determined cost assumption could be valid. However, in many cases, the investment cost is linked to the market price. The assumption  about the investment cost has a crucial influence not only on the value of a real option and its optimal exercise strategy but also on their sensitivity to changes in the input parameters.   
Most papers do not question the effect that the cost assumptions have on the conclusions, ignoring the potential model risk introduced by the cost assumptions, as noted by \citet{S2001}. She demonstrates that the real option value modelled under a fixed cost assumption and a stochastic cost assumption, differ drastically. Similarly, the constant drift assumption in GBM does not allow the modeller to analyse the effect of decreasing/increasing trends in costs. Clearly, if the GBM drift decreases so that expected costs decrease in future, the option value should increase. \citet{F2008}  analyses the effect that a GBM variable cost assumption has on the ROV and concludes that a more appropriate process for variable cost would be mean-reverting.

A fixed cost, where $\alpha=1$ in (\ref{eqn.K}), may be regarded as the strike of an American call option, and the value is derived from the expected utility of a call option pay-off for which the upper part of the terminal wealth distribution above the strike matters. The opposite extreme ($\alpha = 0$) focuses on the lower part of the terminal wealth distribution below the current price $p_0$, where the investment costs are lowest. Although log returns are similar across the whole spectrum under the GBM assumption (\ref{eqn.gbm}), changes in wealth are in absolute terms and are greater in the upper part of the distribution than in the lower part. For this reason an at-the-money fixed-strike assumption yields a greater real option value than the invest-at-market-price assumption.\footnote{However, this property only holds under GBM views for market prices, see Section \ref{sec.boombust} for a counter example under different price processes.} 

We illustrate these properties with an option to purchase a project that has no associated cash flows.  Suppose the current market price of the project is \$1m and the investor believes this will evolve according to (\ref{eqn.gbm}) with $\mu$ and $\sigma$ as specified in Table \ref{tab.costs}. The risk-free lending and borrowing rates are both 5\%. The investment horizon is $T^{\prime} = 5$ years, investors have an exponential utility with risk tolerance $\lambda$, and the initial wealth is \$1m. As $\lambda \rightarrow \infty$ we have a linear utility, giving the option value for a risk-neutral decision maker, and further setting $\mu = r$ gives the RNV solution.  
Decisions are made once per year, so we set $\Delta t = k = 1$. Using (\ref{eqn.jr}) we have $m = 0.03$, $u = 1.259$, $d=0.844$ and $\pi = 0.5$.
 The investment cost takes the general form (\ref{eqn.K}) with $K=\$1$m. We set $g(x) = x$ and compare investment cost at market price ($\alpha=0$) with fixed time 0 cost ($\alpha=1$) and a mix of fixed and variable cost ($\alpha = 0.5$). We also consider $g(x) = x/2$ for variable cost at a fraction (in this case one-half) of the market price, plus a fixed cost if $\alpha > 0$, and set $g(x) = \sqrt{x}$ for a variable cost that increases non-linearly with market price, so the price and cost are not perfectly correlated. When $g(x) = \sqrt{x}$ the variable cost is less than (greater than) the current market price $p^{-}_t$ if $p^{-}_t>\$1$m ($p^{-}_t<\$1$m).

\begin{table}[!h]
\begin{center}~\\
{\footnotesize 
\caption{\small {Exponential utility option values for different risk tolerance $\lambda$ and different investment costs structures, compared with risk-neutral values (linear utility, i.e. $\lambda = \infty$, and $\mu = r$). Here $T^\prime = 5$ years, $\Delta t = k = 1$, $K = p_0 = \$ 1$m, $r =  5\%$.} When $\alpha=1$ (Column 3) the option strike is $K$; when $\alpha=0.5$ the option strike is of the form (\ref{eqn.K}) for the function $g$ specified in row 1 of each section.}
~\\[6pt]
\begin{tabular}{c|c|c|c|cc|cc|cc}
\toprule
    \multicolumn{2}{c|}{$\mu=10\%$} & $g(x)$&-     & \multicolumn{2}{c|}{$x$} & \multicolumn{2}{c|}{$\sqrt{x}$} & \multicolumn{2}{c}{$x/2$} \\
    \cmidrule{3-10}
    \multicolumn{2}{c|}{$\sigma=20\%$} & $\alpha$ &1     & 0.5   & 0     & 0.5   & 0     & 0.5   & 0 \\
\midrule
    \multirow{10}[0]{*}{$\lambda$} & \multirow{2}[0]{*}{0.1} & Value&           47,387  &            29,349  & 0     &            36,260  &            32,222  &          115,185  &          390,296  \\
          &  &Year/State    & 3/$uu$  & 4/$uuu$ & -   & 4/$uuu$ & 4/$uuu$ & 4/$u$   & 4/- \\
\cmidrule{2-10}
          & \multirow{2}[0]{*}{0.5} &  Value&        199,103  &          117,936  &            43,008  &          166,906  &          129,091  &          323,672  &          561,529  \\
          &   &Year/State    & 2/$uu$  & 4/$uu$  & 0   & 4/$uu$  & 4/$uu$  & 4/$u$   & 2/$dd$ \\
\cmidrule{2-10}
          & \multirow{2}[0]{*}{1} &Value&          261,839  &          157,221  &          141,249  &          218,048  &          170,077  &          393,401  &          641,249  \\
          &  &Year/State     & 3/$uuu$ & 3/$uud$ & 0   & 4/$uu$  & 4/$uu$  & 2/$ud$  & 0 \\
\cmidrule{2-10}
          & \multirow{2}[0]{*}{$\infty$} &  Value&        363,789  &          283,179  &          283,179  &          306,855  &          283,179  &          533,179  &          783,179  \\
          &    &Year/State   & 2/$uu$  & 0   & 0   & 1/$u$   & 0   & 0   & 0 \\
\cmidrule{2-10}
          & \multirow{2}[0]{*}{RNV} &   Value&       160,122  &            80,030  & 0     &          124,891  &            89,661  &          270,281  &          499,604  \\
          &    &Year/State   & 3/$uuu$ & 4/$uuu$ & -   & 4/$uuu$ & 4/$uuu$ & 4/$uu$  & 4/- \\
    \bottomrule
\multicolumn{10}{c}{}\\
\toprule
    \multicolumn{2}{c|}{$\mu=15\%$} & $g(x)$  & -     & \multicolumn{2}{c|}{$x$} & \multicolumn{2}{c|}{$\sqrt{x}$} & \multicolumn{2}{c}{$x/2$} \\
    \cmidrule{3-10}
    \multicolumn{2}{c|}{$\sigma=50\%$} & $\alpha$ & 1     & 0.5   & 0     & 0.5   & 0     & 0.5   & 0 \\
    \midrule
    \multirow{10}[0]{*}{$\lambda$} & \multirow{2}[0]{*}{0.1} & Value & 37,277  & 5,374  & 0     & 34,229  & 6,454  & 36,865  & 148,551  \\
          &       & Year/State & 3/$uuu$ & 4/$uuuu$ & -   & 4$/uuu$ & 4/$uuuu$ & 4/$uuu$ & 4/- \\
\cmidrule{2-10}
          & \multirow{2}[0]{*}{0.5} & Value & 151,743  & 36,014  & 3,023  & 125,942  & 82,806  & 181,581  & 341,763  \\
          &       & Year/State & 3/$uuu$ & 4/$uuu$ & 3/$ddd$ & 4/$uuu$ & 4/$uuu$ & 4/$uu$  & 3/$ddd$ \\
\cmidrule{2-10}
          & \multirow{2}[0]{*}{1} & Value & 254,963  & 112,602  & 17,221  & 216,766  & 166,120  & 312,741  & 461,455  \\
          &       & Year/State & 4/$uuu$ & 4/$uuu$ & 2/$dd$  & 4/$uuu$ & 4/$uuu$ & 4/$uu$  & 2/$dd$ \\
\cmidrule{2-10}
          & \multirow{2}[0]{*}{$\infty$} & Value & 823,324  & 608,935  & 608,935  & 720,471  & 624,340  & 879,799  & 1,108,935  \\
          &       & Year/State & 3/$uuu$ & 0   & 0   & 2/$uu$  & 1/$u$   & 1/$u$   & 0 \\
\cmidrule{2-10}
          & \multirow{2}[0]{*}{RNV} & Value & 376,412  & 186,519  & 0     & 306,003  & 235,593  & 359,591  & 485,545  \\
          &       & Year/State & 3/$uuu$ & 4/$uuu$ & -   & 4/$uuu$ & 4/$uuu$ & 4/$uuu$ & 4/- \\
    \bottomrule
    \end{tabular}
\label{tab.costs}}
\end{center}
\vspace{-10pt}
\end{table}

Table \ref{tab.costs} compares the ROVs under each cost assumption, for different decision makers with $\lambda = 0.1, 0.5$ and 1, with the ROV corresponding to the linear utility of a risk-neutral decision maker ($\lambda = \infty$ and $\mu=r$). Under the option value we report the optimal exercise strategy, denoted by year and any market state that it is conditional upon. For instance, $1/d$ denotes invest at time 1 provided the market price moved down between time 0 and 1, and $4/uu$ denotes invest at time 4 provided the market price moved up between time 0 and 1, and again between time 1 and 2, irrespective of later market price moves. Investment at time 0 has no market state, $4/-$ denotes invest in year 4 irrespective of the price state, and never invest is marked simply $-$. 

The option value increases with  $\lambda$, as in \cite{H2007}, because the more risk tolerant the decision maker, the lower the risk premium required to invest. As $\lambda \rightarrow \infty$ the value converges to the risk-neutral (linear utility) value, which is greater than the standard RNV option price  because, in this case,  $\mu > r$. Clearly, the risk-averse ROV can be much less than it is under RNV  especially when risk tolerance is low, or much greater than it is under RNV  especially when risk tolerance is high. Indeed when $\alpha = 0$ and $g(x)=x$, i.e. the investment cost is at the market price, the RNV price is always zero: the CE of a linear utility is the expected value of terminal wealth, and since the discounted price is always a martingale under the risk-neutral measure, CE $=w_0$; so RNV price of an opportunity to invest at market price is always zero. By contrast, the risk-averse decision maker places a positive value on the option in this case, except when $\lambda = 0.1$, when she  would choose not to invest at any market price.\footnote{These observations are not specific to the parameters chosen for Table \ref{tab.costs}. Qualitatively similar properties are evident using other real option parameters, as can be verified by changing the parameters of the downloadable code.} %

The  cost structure also affects the exercise strategy. For a finite $\lambda$  the optimal time to invest is never shorter than for the risk-neutral decision maker, as shown by \cite{H2007}. Further, when $\alpha=0$ and $g(x)=x$, optimal investment is never conditional on a price rise, though it may be conditional on a price fall. Investment also becomes conditional on a price fall but not a rise when $\alpha=0$ and $g(x) = kx$, for $0<k<1$, except that `never invest' is not possible, since the last period pay-off is $x-kx>0$ so an optimal strategy always invests at or before the last period of the option. But when $g(x) = \sqrt{x}$ the last period pay-off $x-\sqrt{x}>0$ only when $x>1$, i.e. after price rise, and optimal investment becomes conditional on up moves. Although we only display results for $k=0.5$ a similar conclusion holds for other $k$ with $0<k<1$. When $\alpha = 0.5$ or 1 there is a fixed cost component with an at-the-money call option pay-off which is positive only if the price rises. Thus,  in Table \ref{tab.costs}   any condition on  optimal exercise is for up moves in price. Finally, the RNV approach typically (but not always) gives an optimal time of investment that is  later than the optimal  time for a risk-neutral decision maker. 
\vspace{-5pt}
\subsection{Exercise Opportunities}
\vspace{-5pt}
We find no previous research which examines the effect of introducing infrequent exercise opportunities. Whether time is discrete or continuous, real option models have previously always assumed that decisions made at time 0 for optimal investment at some time $t\ge0$ can always be implemented at that time $t$. In this section we show how our general framework can model this feature by designating a sub-set of special nodes in the tree as the only opportunities for exercising the investment.  The ROV should not decrease when there is more flexibility in the timing of investments. If the decision never changes as a result of including more or less decision nodes, the option value will remain unchanged. Otherwise, the ROV increases as more decisions are allowed. Having fewer nodes for exercise opportunities places more constraints on the investment opportunity so the project will become less attractive to any decision maker.

\begin{table}[h!]
\begin{center}
\caption{\small {Effect of frequency of exercise opportunities on real option value.  $w_0=p_0 = \$1m, T^{\prime} = 5\mbox{yrs}, \Delta t = 1/12, T=T^{\prime}-k\Delta t, r = 5\%, \mu = 15\%, \sigma = 50\%.$ Exponential utility,  different $\lambda$ and $k$.}}
{\footnotesize $\alpha=0$: invest at market price\\ [1.5ex]
\begin{tabular}{c|c|cccccc}
\toprule
\multicolumn{2}{c|}{$\lambda$}&0.2&0.4&0.6&0.8&1&$\infty$\\
\midrule
\multirow{4}{*}{$k$}&12&109.5&1,132&3,712&8,197&14,480& 645,167\\
&6& 142.2&1,344&4,309&9,212&16,089&645,167\\
&3& 163.9&1,471&4,618&9,803&17,022&645,167\\
&1&176.9&1,553&4,828&10,185&17,624&645,167\\
\bottomrule
\end{tabular}\\
\vspace{0.4cm}
$\alpha=1$: fixed strike with present value \$1m\\
\begin{tabular}{c|c|cccccc}
\addlinespace
\toprule
\multicolumn{2}{c|}{$\lambda$}&0.2&0.4&0.6&0.8&1&$\infty$\\
\midrule
\multirow{4}{*}{$k$}&12&49,385&115,086&174,731&223,201&263,614&881,419\\
&6&71,365&144,638&204,078&252,243&292,632&908,333\\
&3& 86,062&157,289&214,375&263,019&303,834&919,322\\
&1&93,115&166,450&225,568&273,619&313,888&926,058\\
\bottomrule
\end{tabular}}
\label{tab.k}  
\vspace{-10pt}
\end{center}
\end{table}
 
The results in Table \ref{tab.k}  quantify this effect.\footnote{It is important that the trees are nested, i.e. no new decision nodes are inserted as their number decreases, because only in this way does reducing the number of nodes capture the effect of placing additional constraints on decision opportunities.} We again consider  an opportunity to invest in a project with no associated cash flows and current market price $\$1$m. The decision maker has an exponential utility and the real option is characterised by the parameter values given in the legend to the table. Thus, there are 60 monthly steps in the binomial tree for the market price, we place decision nodes every $k$ steps in the backward induction algorithm (\ref{eqn.backind}), and the last exercise opportunity is at $T=T^{\prime}-k\Delta t$. For instance, if $k = 12$ the  nodes occur only once per year and the last exercise opportunity  is taken at the fourth year. So that the four decision trees are nested the values considered for $k$ are $12, 6, 3$ and $1$ representing decision evaluations once per year and once every 6 months, 3 months and 1 month.  For decision makers with different levels of risk tolerance $\lambda$, Table \ref{tab.k} reports results for the two extreme cost cases: the upper part shows the value of the option to invest at market price and the lower part shows the value of the option to invest at a fixed cost $K=\$1$m. In both cases the option values increase when more decision nodes occur in the tree, i.e. as $k$ decreases.

\vspace{-5pt}
\subsection{Risk Tolerance}\label{sec.HARA}
\vspace{-5pt}
ROVs for risk-averse decision makers always increase with risk tolerance, but they can be greater than or less than the RNV option price, depending on the decision maker's  risk tolerance and their expected return on an investment. Two recent papers examine the effect of changing risk tolerance on the value of a real option. 	\citet*{CRS2011} employ a one-factor setting where investment costs consist entirely of risky cash flows, so the marginal cost of investing will increase with risk aversion. Assuming a GBM project value with continuous investment opportunities for a decision maker with HARA utility, the authors show that the marginal benefit of waiting to invest decreases relatively more than costs, as risk aversion increases. As a result, the marginal utility of the pay-off increases, thereby increasing the incentive to postpone investment and the ROV decreases as risk aversion increases. Using a CRRA utility, but in a two-factor GBM setting, 	\citet*{CKS2017} confirm this result. They introduce a model where  the decision-maker's risk attitude depends on her subjective discount rate for the expected utility of the option pay-off, with a lower discount rate being associated with more risk-averse behaviour. As the investor's risk tolerance increases so also does the ROV and the later the optimal timing of the investment.

\begin{figure}[h!]
	\caption{\small {Comparison of invest option values under exponential, logarithmic, power and hyperbolic utilities as a function of risk tolerance.
			Real option values on the vertical scale have been multiplied by 1000 for clarity. Parameters are as in (\ref{eqn.params}).}}
	\centering
	\includegraphics[width=1\textwidth]{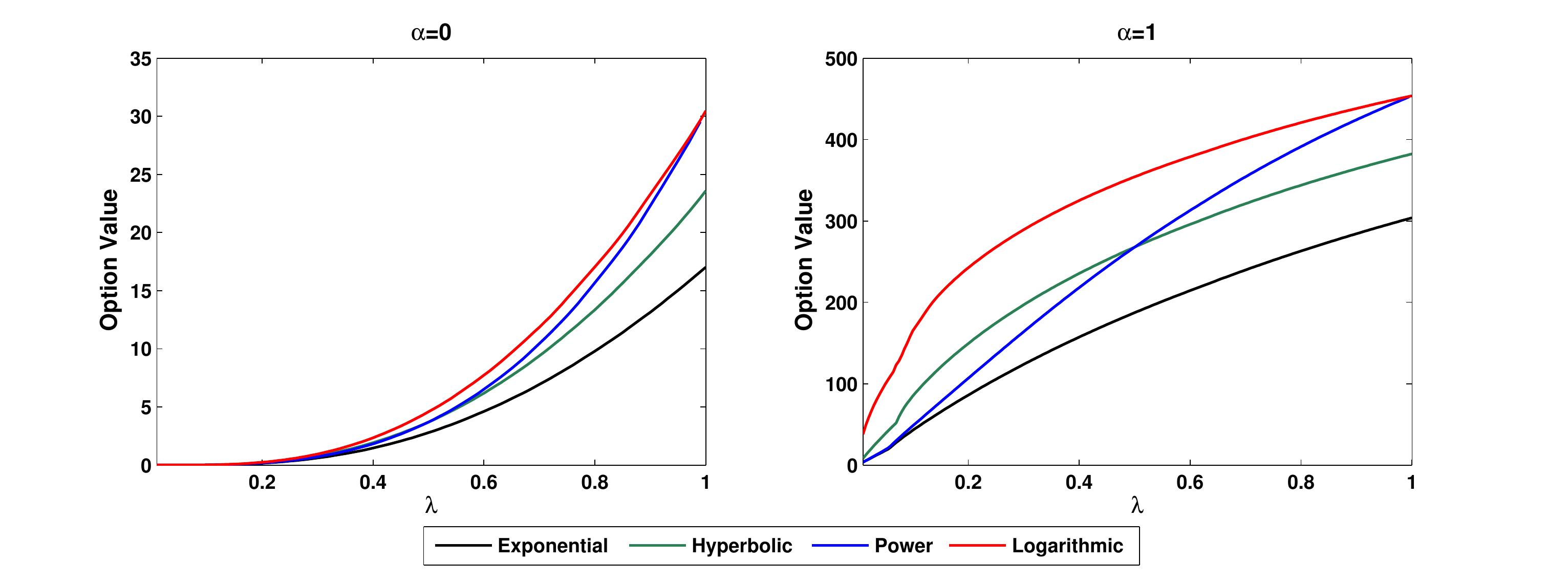}
	\label{fig.comp}
	\vspace{-20pt}
\end{figure}

This section generalises the above result to the HARA utility case to ask how the assumption made about the rate change of risk tolerance with wealth influences the ROV. To this end we assume the local relative risk tolerance coefficient is $\lambda$ at time $0$ and that it increases with wealth linearly at rate $\eta$. To benchmark HARA utility ROVs against exponential utility ROVs, which have constant absolute risk tolerance $\lambda$, we set $w_0=\$1$m. Thus, $\lambda w_0=\lambda$, i.e. the absolute risk tolerance is the same for all utilities at the time of the decision. Figure \ref{fig.comp} considers the options to invest (a) at market price and (b) at a fixed time 0 cost equal to the current market price. We suppose the current market price is $p_0=\$1$m and the other real option characteristics are:\vspace{-10pt}
\begin{equation}\label{eqn.params}
 T^{\prime} = 5\mbox{yrs}, K = \$1\mbox{m}, g(x)=x, \Delta t = 1/12, k =  3, r = 5\%, \mu = 15\%, \sigma = 50\%.\vspace{-.1cm}
\end{equation}

The lines in each figure depict the exponential, logarithmic, hyperbolic and power utility ROVs as a function of $\lambda$, with $0<\lambda \le 1$. For typical value of risk tolerance ($0<\lambda<1$) the exponential and logarithmic utility values provide lower and upper ROV bounds, respectively and the ROVs derived from other HARA utilities lie between these bounds. At very high risk tolerance, i.e. $\lambda > 1$, hyperbolic utility values still lie between the exponential and logarithmic values, but the power utility values exceed the logarithmic values, and as $\lambda$ increases further the power values can become very large indeed because the risk tolerance increases extremely rapidly with wealth. 

Because of the boundary in (\ref{eqn.HARA}) HARA utilities are not always well-behaved, 
unlike exponential utilities. But we have shown that exponential utility values are too low when the decision maker's risk tolerance increases with wealth, which is a more realistic assumption than CARA. For typical values of risk tolerance, power utilities produce reliable ROVs,  but with very high risk tolerance logarithmic or hyperbolic utility representations maybe more appropriate, the former giving ROVs that are greater than the latter. Note that these results also show that we can use exponential and logarithmic utility values as lower and upper bounds for other HARA utility values, within the  range  $0<\lambda<1$.

\vspace{-5pt}
\subsection{Relative Size}\label{sec.size}
\vspace{-5pt}
The larger the project  relative to the decision-maker's initial wealth $w_0$, the higher the current project price and (for fixed $\mu$ and $\sigma$ in the GBM) the more variable the terminal change in wealth. Hence, if the decision-maker is very risk averse, the ROV is likely to decrease as the project size increases (relative to wealth). However, if the decision-maker has very high risk tolerance the opposite could occur, i.e. ROV might actually increase with project size, relative to inital wealth. 

We have found no previous investigation into these questions and how the answers relate to the structure of the investment costs. The closest work is in	\citet*{CKS2017} who argue that, for CRRA utilities with fixed investment costs, ROVs will increase as the size of the project relative to initial wealth decreases. Note that a similar effect can be present for CARA utilities: even though the exponential utility value is independent of initial wealth, as previously noted in Section \ref{sec.utility}, CARA ROVs still depend on the volatility of the option pay-off, and this increases with the size of the investment in the project.

In this section we use our model to quantify the effects of project size by fixing initial wealth $w_0$ at \$1m and supposing $p_0$ is either \$0.1m, \$1m or 10\$m,  keeping all other real option characteristics fixed, at the  values (\ref{eqn.params}). We only depict  exponential and logarithmic ROVs because, from the previous section, these are approximate lower and upper bounds for HARA utilities.

\begin{figure}[h!]
\caption{\small{Real option values under exponential and logarithmic utilities as a function of risk tolerance $\lambda$, $T^\prime=5$, $\Delta t=1/12$, $k=3$,  $r=\tilde r=5\%$, $\mu=15\%$, $\sigma=50\%$, $K=p_0=\$0.1,\, \$1,\, \$10 \mbox{m}$, $w_0=1\mbox{m}$, $0.1\leq\lambda\leq 100$, ROV in \$m. Both axes  in $\log_{10}$ scale.}}
\centering
\includegraphics[width=1\textwidth]{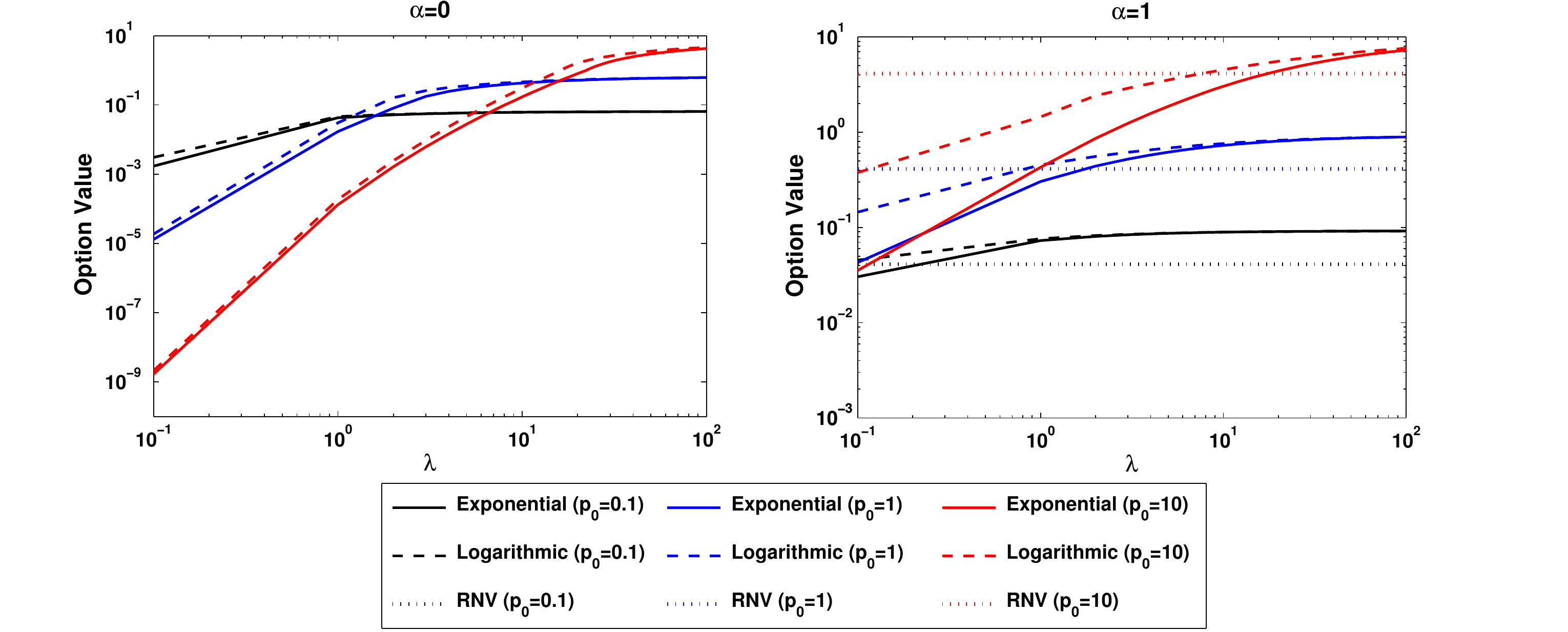}
\label{fig.logexpcomp}
\vspace{-20pt}
\end{figure}

Figure \ref{fig.logexpcomp} displays the ROV for different values of the local risk tolerance $\lambda$ between $0.1$ and $100$, both axes being drawn on a base 10 log scale. We go beyond the range  $0<\lambda<1$ to examine the convergence of ROVs as  $\lambda \rightarrow \infty$, for comparison with the RNV case.  We depict ROVs for  $\alpha = 0$ (invest at market price) on the left, and $\alpha = 1$ (fixed ATM strike at $p_0$) on the right. The values for the largest investment ($p_0=10\,w_0$) are represented by red lines, those for the smallest investment ($p_0=0.1\, w_0$) are in black and those for the medium investment  ($p_0=w_0$) are blue. In each case as $\lambda \rightarrow \infty$, the ROV converges to the value obtained for a risk-neutral investor, which increases with $p_0$. 

In both figures, the relative size of the investment in the project governs the sensitivity of the ROV to $\lambda$, with larger investments having ROVs that are more sensitive to risk tolerance. Consider the left-hand figure ($\alpha = 0$) where the option is to invest at market price ($K_t = p^-_t$) and hence the RNV option price is zero. Given that the drift and volatility of the GBM process driving the market price dynamics is not affected by $p_0$, all decision makers with $0<\lambda<1$ will prefer the smaller investment to the larger ones, but as risk tolerance increases the larger investments become more attractive than the smaller ones. Also note that the parameter $\eta$, i.e. the rate at which risk tolerance grows with wealth, has much lesser effect on the ROV since exponential and logarithmic values remain close to each other throughout. As $\lambda$ increases the ROVs converge towards the risk-neutral subjective value, which is non-zero (because $\mu \ne r$) and which increases with project size. 

By contrast, in the right-hand Figure \ref{fig.logexpcomp} where the costs $K_t$ are fixed at \$1m (i.e. $\alpha = 1$) the RNV, marked by the  dotted horizontal line, is non-zero. It is less than the risk-neutral subjective value to which the ROV converges as $\lambda \rightarrow \infty$, simply because we have chosen $\mu > r$ in this case; if we had set $\mu < r$ the RNV would exceed the asymptotic value in the figure, irrespective of the investment size. 

As we move from a variable to a fixed cost structure the main difference is that  larger investments are always ranked highest. The exponential ROV (solid lines) are now well below the logarithmic ROV (dashed lines) showing that assumptions about $\eta$ can be important,  especially for a highly risk-averse decision maker facing the fixed-strike option to invest in a high-priced project. In general, risk-neutral decision makers attribute higher ranks to larger investments; risk-averse decision makers also tend to prefer  larger fixed-strike investments; but when an investment is at market price, smaller investments are ranked more highly even by investors with very low values risk-aversion.

\section{Model Risk in Price Process  Dynamics }\label{sec.process}
\vspace{-10pt}
\citet{S2001} introduces a model with three state variables -- price, cost, and reserves, finding that the dynamic models that are selected for these processes greatly affect the ROV. In particular, the option values associated with  non-stationary processes are systematically
larger than the comparable stationary values. Also, assuming a GBM for price/cost processes substantially
overvalues the option value relative to the mean-reverting case. However, with two notable exceptions (discussed in more detail below) all the real-options literature which examines the sensitivity of real options to assumptions made about the price process employs the risk-neutral paradigm. 

 Section \ref{sec.musig}, examines how the parameters of a GBM price process affect the ROV. This problem has been analysed in many previous papers, albeit within a particular real-world application where other basic assumptions are made -- typically RNV. So we set our more general results, based on HARA utilities, in the context of this research.\footnote{The decision maker has subjective values for $\mu$ and $\sigma$, but these may be highly uncertain if she lacks confidence in her views. Indeed,  while the decision maker can lose her patience of waiting which leads to an increase in the discount rate applied to the expected utility of the option pay-off it is perhaps more likely that she is highly uncertain about her views, and subsequently adjusts her subjective values for $\mu$ and $\sigma$. In this case, it will be important to examine the sensitivity of ROVs to the expected returns and risks of the project.}   Next, in Section \ref{sec.boombust}  we consider an option to invest in a project when the decision maker's views are captured by a `boom-bust' process  (\ref{eqn.bb}) where a long period of positive price momentum could be followed by a crash. The question of model risk has not previously been analysed in this context.  Finally, in Section \ref{sec.mr} we analyse real options for decision makers who believe in a mean-reverting price process (\ref{eq.OU}). There is some previous literature on this extension, reviewed below, but all the papers that we can find are based on RNV. 

\vspace{-5pt}
\subsection{Risk Sensitivity}\label{sec.musig}
\vspace{-5pt}

ROVs may decrease or increase with volatility, depending on the cost structure of the project and the decision maker's risk preferences. Much prior work, based on a variety of different assumptions for the underlying price process, costs and the valuation measure, finds that ROVs increase with volatility.\footnote{A selection of recent literature that supports this relationship includes: \citet*{F2015}, \citet*{CRS2011}, \citet*{BFJLR2008} and \citet*{AS2004} -- these assume a GBM price process; \citet*{MCP2018} and \citet*{MCCC2011} -- these papers set a mean-reverting price process; Regarding costs, of course RNV option prices always increase with volatility when these are fixed -- but more generally, \citet*{MCP2018}, \citet*{F2015}, \citet*{CRS2011} and \citet*{AS2004} assume with a fixed or deterministic investment cost, whilst \citet*{MCCC2011} and \citet*{BFJLR2008} set the investment cost to be stochastic; for the valuation measure \citet*{MCP2018}, \citet*{F2015}, \citet*{BFJLR2008} and most of the other papers cited here value the  option under the risk-neutral measure -- exceptions include \citet*{CRS2011}, who adopt a HARA utility, \citet*{AS2004} consider a concave utility and \citet*{MCCC2011} employ the physical measure with a risk-adjusted discount rate. } 
 Only a few papers challenge this result, arguing  this relationship can be reversed  in some circumstances. For instance, \citet*{AS2004} assume a fixed cost and a concave utility, and argue that the increase in the volatility can decrease the option value of waiting when the expected utility of the option's pay-off is a convex function over time. \citet*{CKS2017} assume a GBM cost and a CRRA utility, and demonstrate that the ROV can decrease with volatility when the discount rate  is low enough. In this section we use our general framework to capture a drastic decrease or increase in the ROV as $\sigma$ increases, thus demonstrating that the sensitivity of the ROV to volatility depends heavily on the model assumptions. 

\begin{figure}[h!]
	\caption{\small {Option values under exponential and logarithmic utilities with $\lambda = 0.2$, as a function of the investor's subjective views on expected return $\mu$ and volatility $\sigma$. ROV value in \$m for $p_0 = \$1\mbox{m},  w_0 = \$10\mbox{m}, r = 5\%, T^{\prime} = 5, \Delta t = 1/12, k = 3$.}}
	\centering
	\includegraphics[width=1\textwidth]{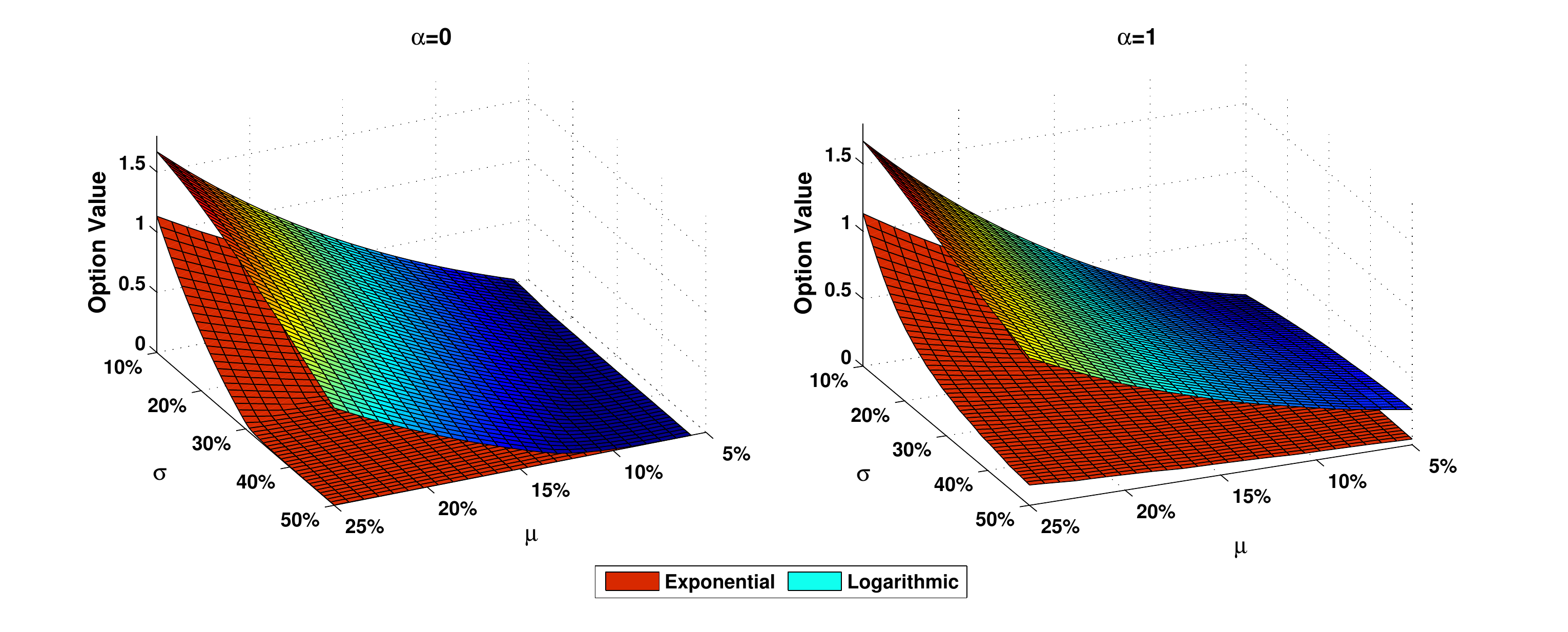}
	\label{fig.sigmu_ivs_dvs}
	\vspace{-20pt}
\end{figure}\normalsize

Figure \ref{fig.sigmu_ivs_dvs} depicts the values of a real option to invest  as a function of the drift and volatility of the GBM market price process, with other parameters fixed as stated in the legend. When $\alpha=0$ (left-hand figure) the option value always decreases as uncertainty increases due to the risk aversion of the decision maker. For high uncertainty ($\sigma$ greater than about 30\%) the exponential utility values are zero: the investment opportunity is valueless because the price would never fall far enough (in the decision maker's opinion) for investment to be profitable. By contrast, the logarithmic utility always yields a positive value when expected return is greater than about 10\%, but again the decision maker becomes more likely to defer investment as the volatility increases. Indeed, the option values are monotonically decreasing in $\sigma$ at every $\mu$, and monotonically increasing with $\mu$ for every $\sigma$. However, the ROV volatility sensitivity is quite different for the fixed-strike option, i.e. for $\alpha = 1$. For instance, the logarithmic ROV can decrease as $\sigma$ increases, when $\mu$ is low. When $\mu = r$ the logarithmic ROV increases monotonically with $\sigma$, as in the RNV case.

The sensitivity of the option value to both $\mu$ and $\sigma$ also decreases as risk tolerance increases. To illustrate this we give a simple numerical example, for the case $\alpha = 0$ and an exponential utility. Suppose $\mu = 20\%$. If $\lambda = 0.2$ the option value is $\$1,148$ when $\sigma = 40\%$ and $\$241,868$  (almost 211 times larger) when $\sigma = 20\%$. When $\lambda = 0.8$ the option value is $\$167,716$ when $\sigma = 40\%$; but now when $\sigma = 20\%$ it only increases by a multiple of about $4$, to $\$713,812$. Similarly, fixing $\sigma = 20\%$ but now decreasing $\mu$ from $20\%$ to $10\%$ the value changes from $\$241,867$ to $\$109$ ($2,210$ times smaller) for $\lambda = 0.2$, but from $\$713,812$ to $\$113,959$ (only about $6$ times smaller) when $\lambda = 0.8$. Hence, the option value's sensitivities to  $\mu$ and $\sigma$ are much greater for low levels of risk tolerance. Similar, but less pronounced, effects are present with the logarithmic and other HARA utilities, as can be verified using the downloadable code.

\vspace{-5pt}
\subsection{Momentum }\label{sec.boombust}
\vspace{-5pt}
Many  markets are subject to booms and recessions, or periodically collapsing bubbles. To name  a few examples:  the price of gold -- which underpins associated mining projects -- recently reached a peak in 2016, since when it has been generally declining; surges in the price of the cryptocurrency Bitcoin  have caused the value of projects based on Initial Coin offering (ICOs) to rise and fall over long periods  during the last few years; and the real-estate market has experienced several booms and busts since the second world war.

Clearly, when the investment horizon is many years away, a decision maker may wish to take account of long periods of upward and/or downward trending markets in her views about expected returns. For illustration, consider a simple boom-bust scenario over a 10-year horizon: the expected return is negative, $\mu_1 < 0$ for the first $n$ years and positive, $\mu_2 > 0$ for the remaining $10-n$ years. Deriving approximate parameter values  from observations on real estate values,\footnote{The average annualised return computed from monthly data on the Vanguard REIT exchanged-traded fund (VQN) from January 2005 to December 2006 was $21\%$ with a volatility of $15\%$. However, from January 2007 to December 2009 the property market crashed, and the VQN had an average annualised return of $-13\%$ with a volatility of $58\%$. But from January 2010 to June 2011 its average annualised return was $22\%$ with volatility $24\%$.}   we set $\mu_1 = -10\%, \sigma_1=50\%$ and $\mu_2 = 10\%, \sigma_2=30\%$. We suppose decisions are taken every quarter with $\Delta t = 0.25$ and set $r = 5\%$ in the price evolution tree. The ROVs given in Table \ref{tab.timevarying} are for decision makers having exponential utility, with varying levels of risk tolerance between $0.2$ and $1$ and the recession is believed to last $n = 0, 2, 4, 6, 8$ or $10$ years.\footnote{Results for the divest option, or for the invest option under different utilities are not presented, for brevity, but are available on request. The qualitative conclusions are similar.}

\begin{table}[h!]
\begin{center}
\caption{\small {Effect of a time-varying drift for the market price, with downward trending price for first $n$ years followed by upward trending price for remaining $10-n$ years.  Exponential utility with different levels of risk tolerance, with $\lambda = \infty$ corresponding to the risk-neutral (linear utility) value. Real option values in bold are the maximum values, for given $\lambda$. $p_0=\$1\mbox{ million}$, $w_0=\$1\mbox{ million}$, $T^\prime=10$, $\Delta t=0.25$, $k=1$, $T=T^\prime-\Delta t$, $r=5\%$, $\mu_1=-10\%$, $\mu_2=10\%$, $\sigma_1=50\%$ and $\sigma_2=20\%$.}}
{\footnotesize $\alpha=0$\\ [1.5ex]
\begin{tabular}{c|c|cccccc}
\toprule
\multicolumn{2}{c|}{$n$}&0&2&4&6&8&10\\
\midrule
\multirow{4}{*}{$\lambda$}&0.2&866&382,918&{\bf 727,922}&438,482&174,637&0\\
&0.8&211,457&1,230,035&{\bf 2,297,909}&1,148,695&370,354&0\\
&1.0&266,752&1,349,295&{\bf 2,552,434}&1,300,724&406,567&0\\
&$\infty$& 648,173&2,234,361&4,777,950&{\bf 5,308,672}&1,045,800&0\\
\bottomrule
\end{tabular}

\medskip
$\alpha=1$\\ [1.5ex]
\begin{tabular}{c|c|cccccc}
\toprule
\multicolumn{2}{c|}{$n$}&0&2&4&6&8&10\\
\midrule
\multirow{4}{*}{$\lambda$}&0.2&164,721  &391,060  &{\bf 552,578}  &194,371  &51,040  &5,022  \\
&0.8&392,866  &961,279  &{\bf 1,704,007}  & 692,529  &159,663  &14,943  \\
&1.0&432,280  &1,064,416  &{\bf 1,938,15}7  & 821,733  &186,257  &17,191  \\
&$\infty$&740,493  &1,989,551  &4,339,659  &{\bf 4,860,376}  &841,706  &50,787  \\
\bottomrule
\end{tabular}}
\label{tab.timevarying}\vspace{-10pt}
\end{center}
\end{table}\normalsize

 When $n=0$ or $n=10$ we have a standard GBM price process, so for any given $\lambda$ the project in the upper of the table, with the invest-at-market-price option ($\alpha= 0$) has a lower ROV than the ROV for the project with the fixed-strike option ($\alpha=1$) in the lower part of the table. However, for intermediate values of $n$ the invest-at-market-price option often has a higher value than the fixed-strike option. This ordering becomes more pronounced as $n$ increases, because when $\alpha=0$ the investment cost decreases with $n$, and if the price rises after investment the profits will be greater than for the fixed-cost option. 
Irrespective of cost structure, in this example the preferred investments are those for which boom and bust periods are of roughly equal length. 

\vspace{-5pt}
\subsection{Mean Reversion}\label{sec.mr}
\vspace{-5pt}
Some earlier research analyses investment  real options based on mean-reversion in the project value process but, like \citet{S2001}, these papers assume risks can be perfectly hedged and therefore employ RNV: \citet*{MCP2018} consider  a discrete time  model with deterministic costs where the underlying commodity price is mean-reverting; and in a two-factor continuous-time setting with mean-reverting project value and investment costs, \citet*{JDSZ2013} show that the investment threshold depends on the ratio between project value and cost.
 
\begin{figure}[h!]
\caption{ \small {Comparison of option values under HARA utilities with respect to mean-reversion rate $\psi$. $w_0=\$1\mbox{m}$, $r=5\%$, $k=1$ and $\lambda=0.4$  $T^\prime=10$, $\Delta t=1/4$, $K=p_0=\$1\mbox{m}$, $T=T^\prime-\Delta t$, $\sigma=40\%$. Characteristic time to mean-revert $\phi=\Delta t/\psi$ in years, e.g. with $\Delta t = 1/4$ then $\psi=0.02 \rightarrow \phi = 12.5$yrs, $\psi=0.1 \rightarrow \phi = 2.5$yrs.}}
\centering
\includegraphics[width=1\textwidth]{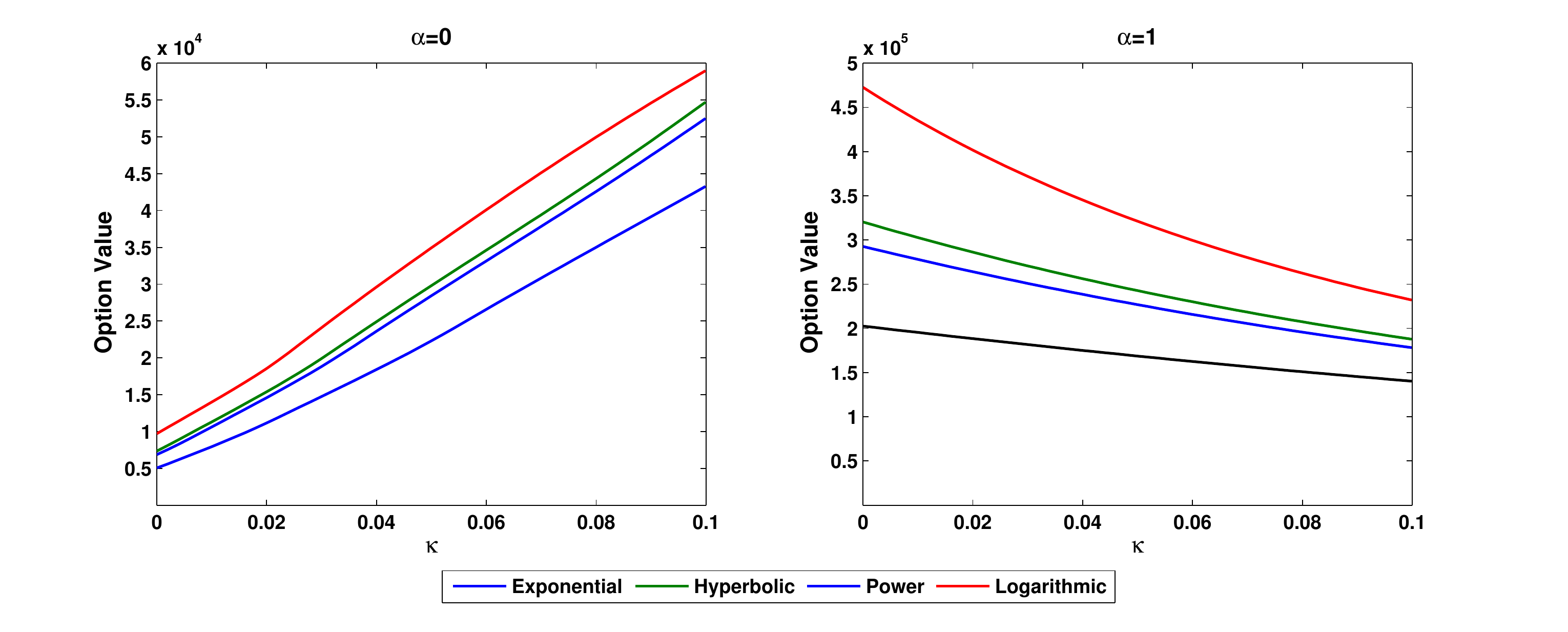}
\label{fig.meanreverting}
\vspace{-20pt}
\end{figure}

In the risk-neutral world it is well-known that an option value will decrease as the mean-reversion becomes stronger; in a sense, increasing the speed of mean-reversion is similar to decreasing volatility. Is this also the case when we apply utility valuation for a risk-averse investor? We can use our model to answer this question by examining how a risk-averse decision-maker's belief in price mean-reversion influences her ROV. 
To this end we assume the price of the project follows the OU process (\ref{eq.OU}) with $\bar p = p_0$. We employ the NR parameterisation (\ref{eq.OUdiscrete}), allowing the speed of mean reversion $\psi$ to vary between $0$ and $0.1$, the case $\psi = 0$ corresponding to GBM and $\psi=0.1$ giving the fastest characteristic time to mean revert of $10$ time-steps (so, assuming these are quarterly, this represents $2.5$ years).\footnote{Lower values for $\psi$ have slower mean-reversion, e.g.  $\psi=0.02$ corresponds to a characteristic time to mean revert of $\phi=0.02^{-1}/4 = 12.5$ years if time-steps are quarterly.} The other parameters are fixed, as stated in the legend to Figure \ref{fig.meanreverting}, which displays the real option values for $\alpha = 0$ and $\alpha= 1$ as a function of $\psi$. 

Increasing the speed of mean-reversion has a similar effect to decreasing volatility. Hence, fixed-cost option values can decrease with $\psi$, as they do on the right graph of Figure \ref{fig.meanreverting} ($\alpha = 1$) especially when the decision maker has logarithmic utility. By contrast, the invest-at-market-price ($\alpha = 0$) option displays values that increase with $\psi$. Note that for fixed $\psi$ these values could \textit{in}crease with $\sigma$, due to the positive effect of $\sigma$ in the local drift (\ref{eq.NRmu}).\footnote{This effect is only evident for values of $\psi$ below a certain bound, depending on the utility function and other real option parameters. For the parameter choice of Figure \ref{fig.meanreverting} the invest-at-market-price option values have their usual negative sensitivity to $\sigma$ once $\psi$ exceeds approximately $2$, where the characteristic time to mean revert is $1/8$th of a year or less. Detailed results are not reported for lack of space, but are available from the authors on request.}

Now consider a risk-neutral decision maker who wishes to rank two  investment opportunities, A and B. Both have mean-reverting price processes but different $\psi$ and $\sigma$: project A has a relatively rapid mean-reversion in its price ($\psi=1/10$, $\phi = 2.5$ years) with a low volatility ($\sigma = 20\%$) and project B has a relatively slow mean-reversion in its price ($\psi=1/40$, $\phi = 10$ years) with a higher volatility ($\sigma = 40\%$). Such an investor would have a great preference for project B since she  is indifferent to the high price risk and has regard only for the local drift. Given the slow rate of mean-reversion, she  sees only the possibility of a sharp fall in price -- at which point she  would invest  -- followed by a long upward trend in its price. Indeed, assuming $p_0=\bar p = \$1$m the risk-neutral value of option B is \$797,486, but the corresponding value for option A is only \$98,077.

\begin{table}[!ht]
	\begin{center}~\\
\caption{\small {Comparison of invest option values for project A ($\psi=1/10$, $\sigma=20\%$) and project B ($\psi=1/40$, $\sigma=40\%$) with $\lambda=0.4$ or $0.8$ . All other parameter values are the same as in Figure \ref{fig.meanreverting}. For each utility we highlight the preferred project in bold.}}\label{tab.meanreverting}
~\\
{\footnotesize \begin{tabular}{c|cc|cc|cc}
\toprule
$\lambda$&\multicolumn{2}{c|}{0.4}&\multicolumn{2}{c|}{0.8}&\multicolumn{2}{c}{$\infty$}\\
\toprule
&{\color{red}A}&{\color{blue}B}&{\color{red}A}&{\color{blue}B}&{\color{red}A}&{\color{blue}B}\\
\midrule
Exponential&{\bf 30,421}&12,952&{\bf 52,732}&43,947&\multirow{4}{*}{98,077}&\multirow{4}{*}{\bf 797,486}\\
Hyperbolic&{\bf 33,526}&17,563&55,230&{\bf 58,909}\\
Power&{\bf 33,045}&16,651&56,365&{\bf 68,051}\\
Logarithmic&19,890&{\bf 21,260}&56,986&{\bf 73,131}\\
\bottomrule
\end{tabular}}
\end{center}

\vspace{-8pt}
\end{table}

\clearpage
However, once we relax the risk-neutral assumption the ranking of these two investment opportunities may change, depending on the utility of the decision maker. The real option values for risk-averse decision makers with high risk tolerance ($\lambda = 0.8$) or low risk tolerance ($\lambda = 0.4$) are shown in Table \ref{tab.meanreverting}. In each case the greater option value is depicted in bold. This shows that an investor with a logarithmic utility would also prefer project B, as would an investor with a relatively high risk tolerance ($\lambda=0.8$) and a hyperbolic or a power utility. In all other cases the decision maker would prefer project A.

\section{Summary and Conclusion}\label{sec.conclusion}
\vspace{-10pt}

The value of an investment real option is the net present value that, if received with certainty, would give a risk-averse decision maker the same utility value as the expected utility of the uncertain investment. Such values enable the decision maker to rank  opportunities to invest in alternative projects and the process of valuing the option also specifies an optimal time to exercise. The minimum value of zero applies when the project would not be attractive whatever its future value. 

The special case of RNV, while most commonly employed in the literature, only applies to a real option that is tradable on a secondary market. Therefore, we introduce a general methodology with applications to a wide range of private investment or divestment decisions where project risks are based on a source of uncertainty which cannot be hedged. It permits risk-averse private companies, publicly-funded entities, or individuals to compute a real option value that is completely tailored to the decision maker, based on a very flexible risk preference model (HARA),  where views on the dynamics of the project price process can take several forms and with investment costs that have both fixed and variable components. 

In this paradigm, ROVs can be very different from the risk-neutral price obtained under the standard but (typically) invalid assumption of perfectly-hedgeable risks. This is significant because such decisions can have profound implications for the decision maker's economic welfare. For instance, an individual's investment in housing may represent a major component of her wealth and should not be viewed simply in expected net present value terms, nor should all uncertainties be based on systematic risk because they are largely un-hedgeable. Private companies typically generate returns and risks that have a utility value that is specific to the owners' outlook. Likewise, publicly-funded entities may have objectives that are far removed from wealth maximisation under risk-neutrality.

The real option price that is obtained using standard RNV assumptions can be very much greater than or less than the value that would be found using a more realistic assumptions about investment costs in an incomplete market, and very often the RNV approach would specify a later investment time. When risks cannot be hedged, the risk-averse decision maker's ranking of different real options depends heavily on how her risk tolerance changes with wealth, and so it becomes necessary to employ a HARA utility function in order to capture this feature of model risk. 

We have introduced a very general ROV model and applied it to answer several important questions relating to the model risk in real option models, many of which have not previously been addressed. Our main conclusions can be summarised as follows: \\ \vspace{-10pt}
\begin{enumerate}
	\item [(i)] The assumption about the investment cost, whether it is pre-determined or stochastic, has a significant influence on the real option value. The pre-determined-strike assumption can significantly over-estimate the value of a real option when the more-appropriate assumption is that the investment cost is positively related to the market price, or has both a fixed and variable components.   Fixed-strike options may be more realistic for a decision maker having significant market power, who can use this power to influence the investment cost in her favour. Intuitively,  real options can be more valuable for those decision makers -- less powerful decision makers have to bear more uncertainty in investment costs.  
	\item [(ii)] It is important to account for the flexibility of the decision-making process in real option analysis because the  real option value increases with the frequency of decision opportunities. Hence, due to operational reasons (e.g. infrequent board meetings), in reality, the real option value should be much lower than the standard continuous-time, risk-neutral  value.  
	\item [(iii)] ROVs within the HARA class are bounded above by logarithmic utility values with the (more usual) exponential utility values providing only a lower bound. The option value function is a convex (concave) monotonically increasing function of risk tolerance when investment costs are perfectly correlated (uncorrelated) with the project value.
	\item [(iv)] The value of the project relative to the decision maker's wealth matters a great deal. For instance, the smaller (greater) the risk tolerance of the decision maker, the higher she  ranks the option to invest in a relatively low-priced (high-priced) project, when the project price dynamics follow the same GBM process. Intuitively, the more risk averse, the more the decision maker would favour a small project over a more valuable but costly one, even if two projects are exposed to the same risks. 
	\item [(v)] Under the GBM assumption the sensitivity of a fixed-strike real option to the volatility of the market price may be always positive (as in the RNV approach)  but there are some cases when it may be always negative. This sensitivity increases as risk aversion increases. Our framework is flexible enough to capture both positive and negative volatility effect on the ROV; 
	\item [(vi)] Projects with values which trend strongly over long periods of time tend to be less attractive to risk-averse investors than those with mean-reverting project values. When projects value mean-revert the option value can increase or decrease with risk tolerance, depending on the cost structure.

\end{enumerate}
	
\noindent To aid further research in the area of decision-tree analysis for real options valuation, Matlab code for all the examples can be downloaded by the reader and the parameters may be changed to any reasonable value, set by the user in order to compute values for particular real options that are formulated to model real-world decision problems. This way, any number of specific results on model risk may be generated.
 
 \newpage
\singlespacing
\vspace{-15pt}
\def\newblock{\hskip .11em plus .33em minus .07em}
\bibliographystyle{plainnat}
\bibliography{RoPaper}

\clearpage
\section*{Appendix: Numerical Examples}\label{sec.examples}
\onehalfspacing
\vspace{-10pt}
To fix ideas, here we depict some simple decision trees representing different types of real options. 
In this section we shall only illustrate the simple cost structure \eqref{eqn.K} with $\alpha = 0$ and $g(x)=x$, i.e. the transacted price is the market price of the project. Matlab code is available from the authors for these and the more complex examples included  in the paper. 

First we consider a decision opportunity which provides a concrete example of the zero-correlation case considered by \cite{G2011}, where the opportunity to invest should still have a positive value. A second example shows that the equivalent divest decision also has a positive value. The tables in this section rank  competing 
projects by computing  ROVs under various assumptions: (i) the decision-maker's (subjective) market price for each alternative  follows  the dynamics \eqref{eqn.gbm} with a different (project-specific) expected returns $\mu$ and volatilities $\sigma$. We also consider competing projects with different dividends $\delta$;  (ii) the decision-maker has risk tolerance $\lambda=0.4$, i.e. a local absolute risk aversion of 2.5; and (iii) her initial wealth is equal to the current market price of the project, with  $w_0=p_0=\$1$m; the risk-free interest rate is constant, $r=5\%$ and exercise opportunities occur each time the underlying price moves. We conclude this section by showing how  options with negative cash flows, such as a pharmaceutical research project or land development, are represented in our general framework.

\vspace{-5pt}
\subsection{Positive Cash Flows}\label{sec.rents}
\begin{figure}[h!]
	\caption{\small {Option to invest in a property that pays rents, $x_{{\bf s}(t)}$. $T^{\prime}=3, T = 2, k = 1$. Terminal nodes labelled with $(w-w_0)$, cash flows given by (\ref{eqn.CF}) with with CF = $x$ following decision to invest ($I$), and $0$ otherwise.}}
	\centering
	\includegraphics[scale=0.65]{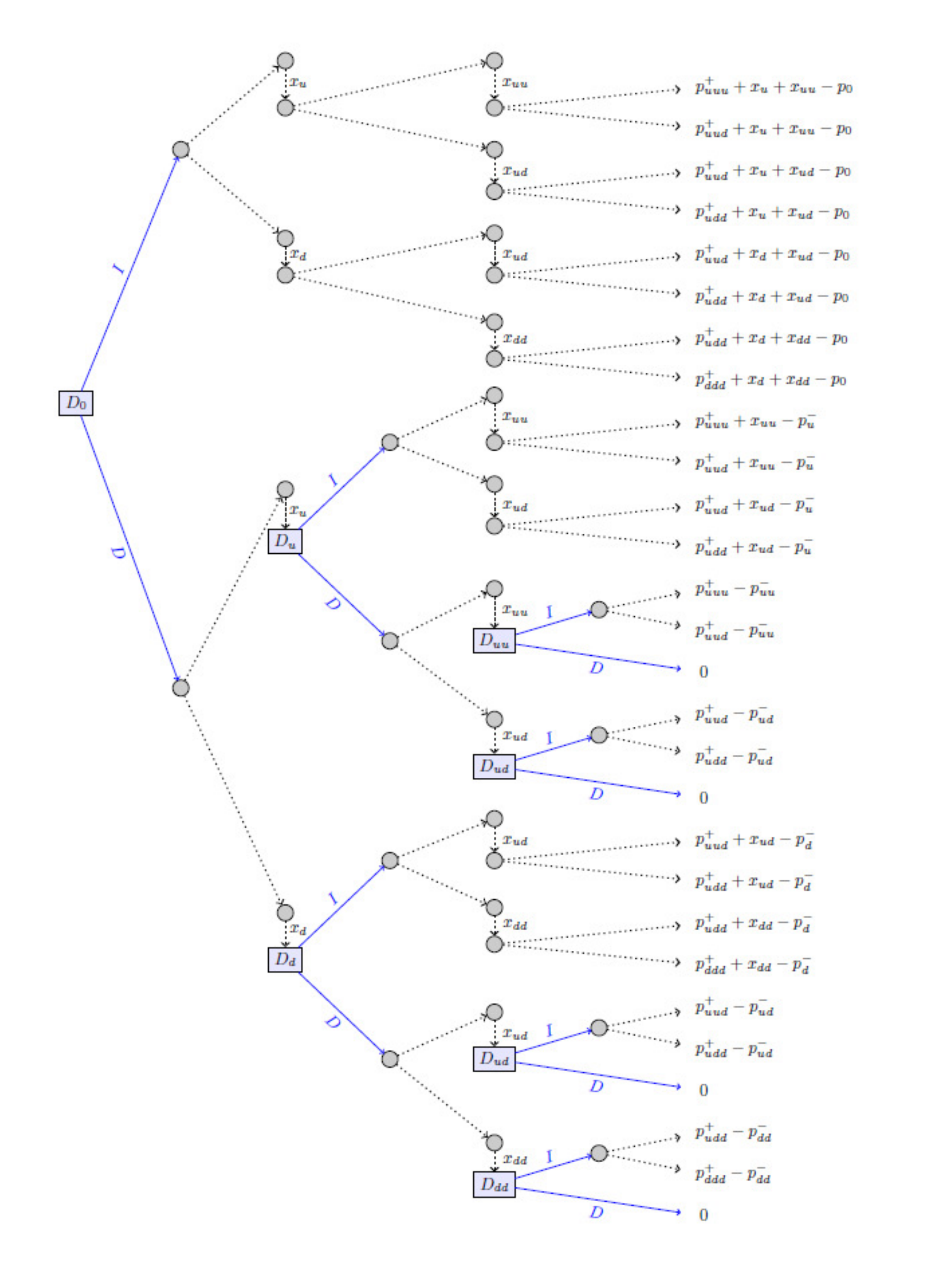}
	\label{fig.buy_rent}
	\vspace{-20pt}
\end{figure}

\begin{figure}[h!]
	\caption{\small {Option to sell a property paying rents, $x_{{\bf s}(t)}$. $T^{\prime}=3, T = 2, k = 1$. Terminal nodes labelled  $\left(w-w_0\right)$, cash-flows are (\ref{eqn.CFdis}) with CF = $x$ if the owner remains invested ($R$), or if the owner sells the property ($S$) the cash-flow is difference between the selling price and initial price.}}
	\centering
	\includegraphics[scale=0.68]{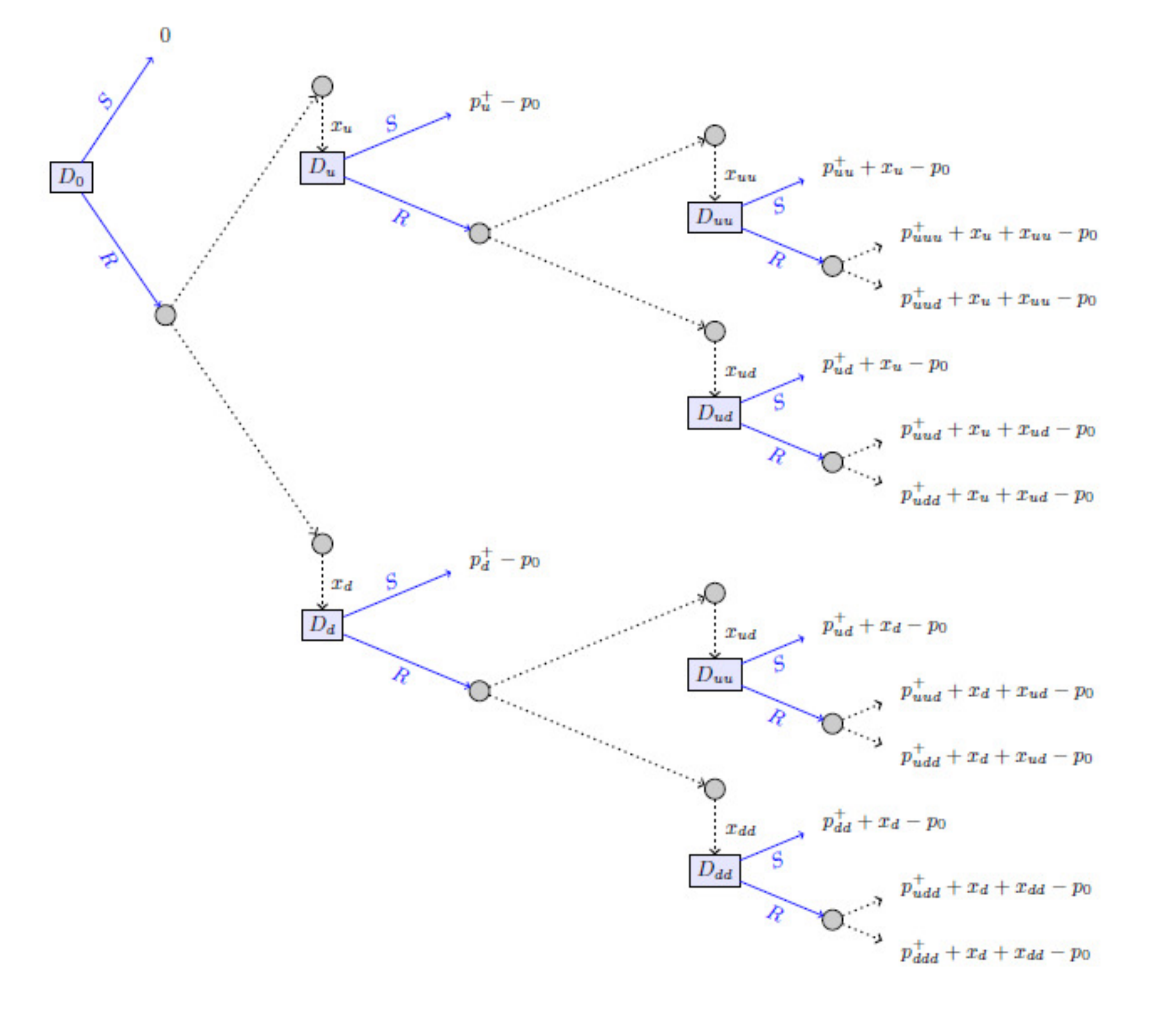}
	\label{fig.sell_rent}
	\vspace{-20pt}
\end{figure}

\vspace{-5pt}
Figures \ref{fig.buy_rent} and \ref{fig.sell_rent} depict two short-horizon decision trees for the invest and divest options on a project that yields positive cash flows. For instance, the project could be the purchase of real-estate with regular positive cash flows, such as a buy-to-let residential or office block, or a car park where fees accrue to its owner for its usage. Rents, denoted $x_{{\bf s}(t)}$ in the trees, are captured using a positive, constant dividend yield defined by (\ref{eqn.div}), to capture rents that only increase/decrease in line with the market price. We suppose cash flows are not re-invested and instead earn the risk-free rate. 
Each time a cash flow is paid the market price jumps down from $p^+_t$ to $p^-_t=p^+_t-x_t$. Between payments the decision maker expects the discounted  market price to grow at rate $\mu - r$, and based on the discretisation (\ref{eqn.jr}) we have  $p^+_{t+1} = up^-_t$ or $p^+_{t+1} = dp^-_t$ with equal probability. The terminal nodes of the tree are associated with the increment in wealth $w - w_0$ where the final wealth $w$ is given (\ref{eqn.CF}) for the option to invest in Figure \ref{fig.buy_rent}, and by (\ref{eqn.CFdis}) for the option to divest (setting  the cash flow equal to the rent in each state) depicted in Figure \ref{fig.sell_rent}.

Table 	\ref{tab.rent} applies the decision trees in Figures \ref{fig.buy_rent} and \ref{fig.sell_rent} to rank the options to buy (Fig. \ref{fig.buy_rent}) and to sell (Fig. \ref{fig.sell_rent}) two different buy-to-let properties. 
In each case rents are paid every six months, and are set at a constant percentage $\delta$ of the market price at the time the rent is paid. But the investor has different views about the future market price and rents on each property, as characterised by different values for $\mu$, $\sigma$ and $\delta$. 

\vspace{10pt}
\begin{table}[h!]
	\caption{\small {Columns 2 and 3 compare the values of 2 real options, each to purchase a buy-to-let property as depicted in Figure \ref{fig.buy_rent}. Columns 4 and 5 compare the values of 2 other options, each to sell a buy-to-let property currently owned, as in Figure \ref{fig.sell_rent}. In each case our decision-maker has $\lambda = 0.4$ and $w_0=\$1$m,  each property has $p_0=\$1$m and $r=5\%$. The decision maker's beliefs about $\mu$, $\sigma$ and $\delta$ depend on the property. For each utility we highlight in bold the preferred location for buying (or selling) the property.}}
	\begin{center}
		{\footnotesize \begin{tabular}{c|cc|cc}
				\toprule
				& \multicolumn{2}{c|}{Invest} & \multicolumn{2}{c}{Divest} \\
				\midrule
				$\sigma$&40\%&25\%&30\%&20\%\\
				$\mu$&15\%&10\%&15\%&10\%\\
				$\delta$&10\%&10\%&20\%&10\%\\
				\midrule
				Exponential& 289   & \bf{316}   & \bf{6,775} & 6,052 \\
				Hyperbolic & \bf{348}   & 339   & \bf{6,610} & 6,103 \\
				Power & \bf{340}   & 336   & \bf{7,283} & 6,296 \\
				Logarithmic & \bf{358}   & 345   & 4,286 & \bf{5,330} \\
				\bottomrule
		\end{tabular}}
		\label{tab.rent}
	\end{center}
	\vspace{-10pt}
\end{table}\normalsize

Similarly, we rank the decisions (depicted in Figure \ref{fig.sell_rent}) to sell two different properties, according to the subjective views on market prices and rents  specified in the second pair of columns in Table \ref{tab.rent}. The divest ROV is the expected utility value of capital gains on the property  plus the expected utility value of the opportunity to divest with the value of the preferred property  marked in bold.

Clearly, the form assumed for the utility function has material consequences for decision making. A decision maker with CARA utility would prefer the option to buy the second property, whilst decision makers with any of the other HARA utilities (and the same risk tolerance at time 0) would prefer to buy the first property. And for the divest real option values of two other properties both currently owned by the decision maker, all decision makers except those with a logarithmic utility would favour selling the first property.

\vspace{-5pt}
\subsection{Negative Cash Flows}\label{sec.devcosts}
\vspace{-5pt}
In the buy-to-develop case there are no cash flows until the  research and development of a pharmaceutical project starts, or the building plot is purchased. Thereafter, these negative cash flows should be accounted for in the market price. Setting a negative, constant dividend yield cannot capture these properties. Instead, the market price following an invest decision is cum-dividend, but prior to this it evolves as in the zero cash-flow case. Another important realism to include in the decision-tree is that the investment horizon $T^{\prime}$ is path dependent now, because it depends on the time of the investment to buy the rights. For instance, it could take 2 years to develop the drug after purchasing the patent rights.

\begin{figure}[h!]
	\caption{\small {Decision tree depicting the option to invest in a project for development. If the rights to development are acquired ($I$) the development takes 2 periods and development costs occur only after the first period. Terminal nodes are associated with the P\&L, $w-w_0$ resulting from the decision.}}
	\centering
	\includegraphics[scale=0.65]{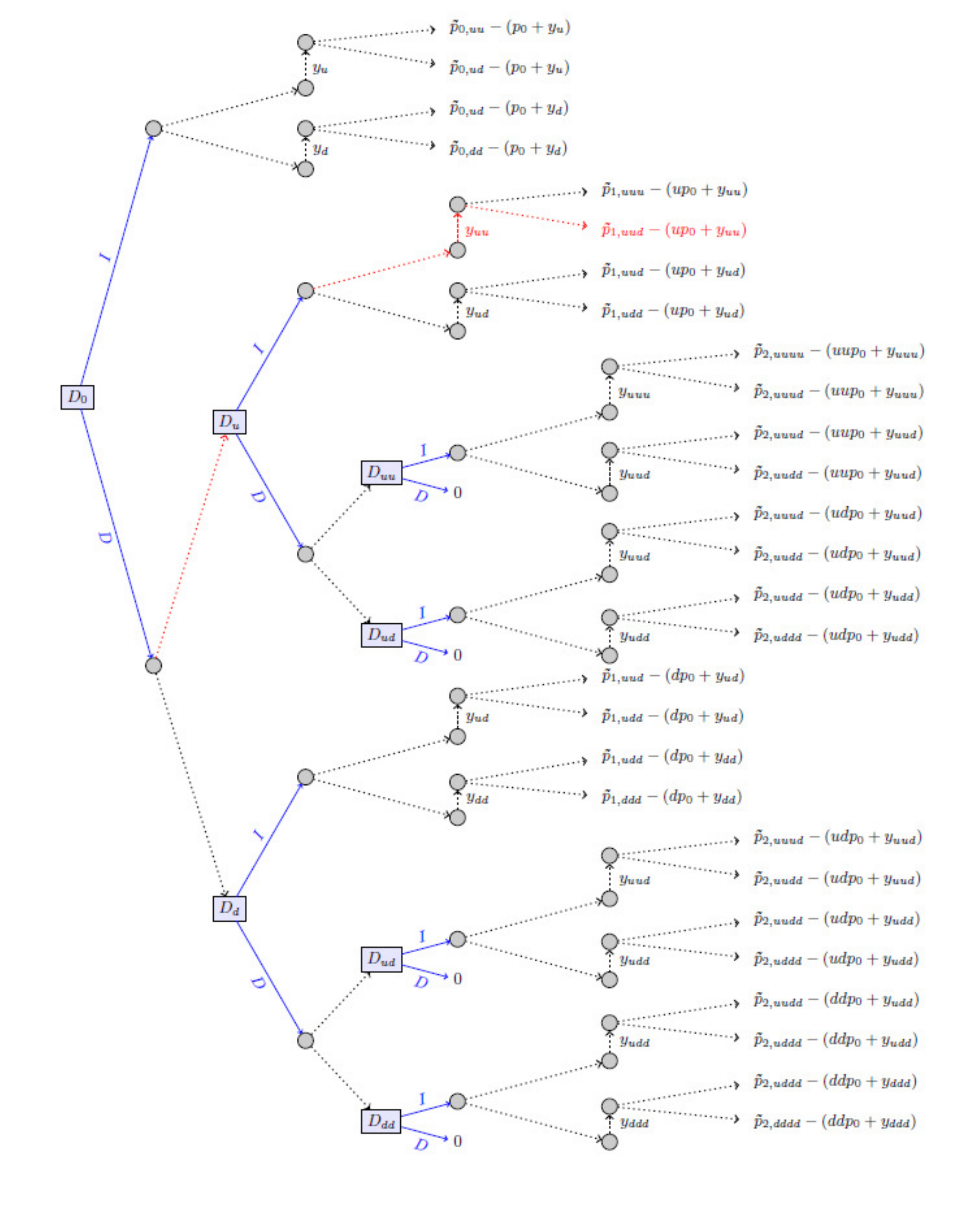}
	\label{fig.devcost}
	\vspace{-20pt}
\end{figure}

A simple  buy-to-develop option is depicted in Figure \ref{fig.devcost}, in which the development cost is $y_{{\bf s}(t)} > 0$ and $\tilde{p}_t$ is the cum-dividend market price. The option maturity $T$ is 2 periods, and so is the development time, so $T^{\prime}$ varies from 2 to 4 periods depending on the time that the project begins. To keep the tree simple we suppose that development costs are paid only once, after 1 period, to allow for planning time after purchasing the rights. 
For example, consider the node labelled $D_u$ that arises if the investor does not purchase the rights at $t = 0$ and subsequently the market price moves up at $t=1$. A decision to invest at this time leads to four possible outcomes.  For instance, following the dotted red lines, if the price moves up again at $t=2$ the development cost at this time is $y_{uu}$, based on the market price of $uup_0$. But if the price subsequently moves down at $t=3$, the terminal value of the property is $\tilde{p}_{1,uud} = d(uu p_0 + y_{uu})$ and the costs are the sum of the price paid for the land and the development cost, i.e. $up_0 + y_{uu}$.

Table \ref{tab.devcost} displays some numerical results for the decision tree in Figure \ref{fig.devcost}, reporting the value of two options to buy-to-develop a project, each with initial market price $\$1$ million and $r = 5\%$, but the options have different $\mu$, $\sigma$ and development costs $\delta$. In each case the development takes one year in total with the costs paid six months after purchase. 
In general, a higher development cost for given $\mu$ and $\sigma$ decreases the buy-to-develop option value, and the decision maker becomes more likely to defer investment until the market price falls. But the option value also increases with $\mu$ and decreases with $\sigma$. We find that option A is preferred by an decision maker with an exponential or logarithmic utility whereas option B is preferred by an decision maker with a hyperbolic or a power utility. Hence, different investors that have identical wealth, share the same initial risk tolerance, and hold the same views about development costs and the evolution of market prices may again rank the values of two development options differently, just because their risk tolerance has different sensitivity to changes in wealth. 

\begin{table}[h]
		\begin{center}~\\
	\caption{\small {Value comparison of real options to buy two different rights for developing a patent for an R\&D project, depicted in Figure \ref{fig.devcost}. In each case our decision-maker has $\lambda = 0.4$ and $w_0=\$1$m,  each project has $p_0=\$1$m and $r=5\%$. The decision maker's views on $\mu$, $\sigma$ and development costs $\delta$ differ for each project, as shown in the table. The preferred option is indicated by the value in bold. }}
~\\
		{\footnotesize \begin{tabular}{c|cc}
				\toprule
				&{\color{red} A}&{\color{blue} B}\\
				\midrule
				$\sigma$&25\%  & 15\% \\
				$\mu$    & 10\%  & 35\% \\
				$\delta$ & 20\%  & 40\% \\
				\midrule
				Exponential & \bf{289}   & 182 \\
				Hyperbolic & 299   & \bf{326} \\
				Power& 315   & \bf{350} \\
				Logarithmic & \bf{41}    & 0 \\
				\bottomrule
		\end{tabular}}
		\label{tab.devcost}
	\end{center}
	\vspace{-10pt}
\end{table}\normalsize

\normalsize
\clearpage

\normalsize

\end{document}